\documentclass[preprint,superscriptaddress,showpacs]{revtex4-1}
\usepackage{epsfig}
\usepackage{graphics,graphicx}
\usepackage{amssymb}
\usepackage{mathrsfs}
\usepackage{amsmath}
\usepackage{verbatim}
\usepackage{hyperref} 
\usepackage{rotating}
\usepackage{float}
\usepackage{setspace}

\begin{document}

\title{Structural behavior and dynamics 
of an anomalous fluid between solvophilic and solvophobic walls:
templating, molding and superdiffusion}  

\date{\today}

\author{Fabio Leoni}\thanks{Present address: School of Mechanical
  Engineering, Tel-Aviv University, Tel-Aviv 69978, Israel}
\affiliation{Departament de Fisica Fonamental, Universitat de
  Barcelona, Mart\'{\i} i Franqu\`es 1, 08028 Barcelona, Spain}
\author{Giancarlo Franzese}
\affiliation{Departament de Fisica Fonamental, Universitat de
  Barcelona, Mart\'{\i} i Franqu\`es 1, 08028 Barcelona, Spain}

\begin{abstract}
Confinement can modify the
dynamics, the thermodynamics and the structural properties 
of liquid water, the prototypical anomalous liquid.  
By considering a general anomalous liquid, suitable for 
globular proteins, colloids or liquid metals, we study by molecular
dynamics simulations 
the effect of a solvophilic structured and a solvophobic unstructured
wall on the phases, the crystal nucleation and the
dynamics of the fluid. 
We find that at low temperatures the large density of the
solvophilic wall induces a high-density, high-energy structure in
the first layer  (``templating'' effect). In turn, the first layer
induces a ``molding'' effect on the second layer determining a
structure with reduced energy  and density, closer to 
the average density of the system. This low-density, low-energy
structure propagates further through the layers by templating effect
and can involve all the existing layers at the lowest temperatures
investigated. Therefore, although the  high-density, high-energy
structure does not self-reproduce further than the first layer, the
structured wall can have a long-range effect thanks to a sequence of
templating, molding and templating effects through the layers.
We find dynamical slowing down of the solvent near the solvophilic
wall but with largely heterogeneous dynamics near the wall due to
superdiffusive liquid veins within a frozen matrix of solvent.  Hence,
the partial freezing of the first hydration layer does not correspond
necessarily to an effective reduction of the channel section in terms
of transport properties. 
\end{abstract}

\maketitle

\section{Introduction}
\label{sec:introduction}

The study of the properties of liquids confined at the nanometer scale
is a topic of high interest for its technological, experimental and
theoretical implications 
\cite{Bellissent95, schoen1998, Mashl2003, Mittal2008, cicero2008,
  De-Virgiliis2008, Mallamace2008,
  Giovambattista09, Mancinelli2009, Gallo2010, Rzysko2010, Han2010, Truskett01,
  delosSantos2011, 
  schnell2011, schnell2012, Nair2012, Paul2012, Ferguson2012, Stewart2012, Krott2013}.
Furthermore, confinement plays an important role in hydrated biological
systems and organic solvents \cite{Schiro2009, Karan2012, Biedermann2013, Franzese2013}.
Structural, thermodynamical and dynamical properties of a liquid can
change near an interface (solid, liquid, etc.) \cite{giovambattista2012}.
When the surface-to-volume ratio is large, at least along one
direction as for the slit pore geometry, the effect of the confining
surfaces has to be taken into account. 
Experiments and simulations on nanoconfined fluids show that molecules
arrange in layers parallel to the surface. The effect becomes stronger
for decreasing temperature or increasing density, until the fluid
eventually solidifies. 
The nature of the solid, as an amorphous or a crystal, can depend on
the interparticle potential and the confinement conditions
\cite{jinesh2008,koga1997,zangi2003,slovak1999}.  
In different cases the role of the interfaces can results in complex
behaviors, e.g. a persisting fluid mono-layer around a spherical
impurity while the rest of the system is in a polycrystal or glassy
state \cite{villeneuve2005}, or persisting amorphous water mono-layer
near an hidrophilic disordered surface \cite{limmer2012}.
In a recent experiment \cite{Kaya2013}, Kaya and coworkers found that
a thin-film water on a $\mbox{BaF}_2(111)$ surface remains in a high
density liquid form for temperatures ranging from ambient (300K) to
supercooled (259K). The result is unexpected because, based on
thermodynamics arguments \cite{sadtchenko2002,foster2000}, Kaya and
coworker would expect that the templating effect of
$\mbox{BaF}_2(111)$ on the structure of the water film should promote
the tetrahedral structure of the low-density liquid.
Moreover, the presence of an interface can promote the heterogeneous
nucleation of the crystal. However, recent experiments and simulations
based on crystallographic analysis of liquids on crystals
\cite{Cox:2012vn}, or on colloidal self-assembly \cite{Chen:2011ve}, show
that the traditional theories need to be revised in these cases.
Confined fluids, under suitable conditions of density and temperature,
can spontaneously develop patterns, e.g. stripes, and different
mesophases or more complex structures \cite{MalPel2003,Glaser2007}. 
The confinement affects also the dynamics of the liquid. It has been
found that near an interface there is a reduction of the local
diffusivity of the liquid \cite{Hansen2013}. 
Computer simulations can help in interpreting the experimental results
for nanoconfined fluids that are difficult to understand
\cite{zhang2011,soper2011,zhang2011b,bai2012,Kaya2013}.
Different numerical approaches based on first principles simulations
can give detailed informations, but are limited by their high
computational cost.   
Classical molecular dynamics of empirical fluid models employ
parameters tested for the bulk case that not necessarily hold in
confinement.
The difficulty to adopt these bulk fluid models to the case of
confinement leads to a variety of simulation results concerning the
aggregation state of the fluid near the surfaces that in principle are
model-dependent. 
It is, therefore, useful to develop coarse-grained models that allow
for analytic calculations
\cite{Ja98,Fra01,Fra02,Sk04,Yan05,Xu2005,Ol06a,Ol06b,FS07,Barraz:2009et}
and more efficient simulations \cite{Gu02}, like isotropic pairwise
core-softened potentials, and that could allow us to better understand
common features of fluids under confinement and the basic mechanisms
of complex phenomena emerging in these systems, like pattern
formation, e.g. stripes and different mesophases. 
This has been confirmed in a recent work \cite{Dotera2014} in which 
the authors claim that the origin of quasi-crystals could be
understood in the context  of a coarse-grained model by the competing
effect of the hard and soft core radius of interacting particles.

Here we focus on the study of nanoconfined anomalous fluids, relevant
for biological and technological applications
\cite{zangi2004,gelb1999}, by means of molecular dynamics simulations
of a system of identical particles interacting through the continuous
shouldered well potential (CSW), an isotropic pairwise core-softened
potential with a repulsive shoulder and an attractive well
\cite{Fr07a,OFNB08}.  
The CSW fluid is confined in a slit pore obtained by a
solvophilic structured wall, and a solvophobic wall with no structure,
as described in Sec.\ref{sec:simulations}.  
The CSW model is suitable for studying globular proteins in solution
\cite{C3SM50220A}, specific colloids
\cite{Lowen1997129,PhysRevE.55.637,PhysRevLett.74.2519}, and liquid
metals \cite{Bryk:2013fk}, and displays water-like anomalies
\cite{Ma05}. 
In particular the CSW reproduces density, diffusion and structure
anomalies following the water hierarchy \cite{Ma05} and
displays a liquid-gas and a liquid-liquid (LL) phase transition, both
ending in critical points \cite{OFNB08}.  

Here we focus on structural and dynamical properties of the CSW
model in confinement and show the role that the characteristic length
scales of the inter-particle potential have in the structuring of the
fluid.


\section{Methods}
\label{sec:methods}

\subsection{The model}
\label{sec:description}

We consider a system of $N$ identical particles interacting by means
of the CSW potential confined between two parallel walls. The CSW
potential is defined as \cite{Fr07a} 
\begin{equation}
U(r)\equiv\dfrac{U_R}{1+\exp(\Delta(r-R_R)/a)}-U_A\exp\left[-\dfrac{(r-R)^2}
  {2\delta_A^2}\right]+\left(\dfrac{a}{r}\right)^{24}   
\end{equation}
where $a$ is the diameter of the particles, $R_A$ and $R_R$ are the
distance of the attractive minimum and the repulsive radius,
respectively, $U_A$ and $U_R$ are the energies of the attractive well
and the repulsive shoulder, respectively, $\delta_A^2$ is the variance
of the Gaussian centered in $R_A$ and $\Delta$ is the parameter which
controls the slope between the shoulder and the well at $R_R$.
The parameters employed are the same as in
Refs.\cite{OFNB08,Fr07a,vilaseca:084507}: $U_R/U_A=2$,
$R^*_R=R_R/a=1.6$, $R_A^*=R_A/a=2$,
$(\delta^*_A)^2=(\delta_A/a)^2=0.1$. In order to reduce the
computational cost, we impose a cutoff for the potential at a distance
$r_c/a=3$.   
In the present simulations we use $\Delta=15$ that allows to better
evidence the anomalies in density, diffusion and structure
\cite{OFNB08,Fr07a}. 


\subsection{Theoretical and simulation details}
\label{sec:simulations}

In our simulations we consider a $NVT$ ensemble system composed of
$N=1024$ particles at fixed temperature $T$ and volume $V$.   
The temperature of the thermal bath $T$ is kept constant by
rescaling the velocity of the particles at each time step by a factor
$(T/{\mathcal T})^{1/2}$, where ${\mathcal T}$ is the instantaneous kinetic
temperature (Allen thermostat) \cite{allen1989}. 
Pressure, temperature, density and diffusion constant are all
expressed in internal units: $P^*\equiv Pa^3/U_A$, $T^*\equiv k_BT/U_A$,
$\rho^*\equiv\rho a^3$, and $D^*\equiv D(m/a^2U_A)^{1/2}$, respectively. 
The equation of motion are integrated by means of the velocity Verlet
method \cite{allen1989}, using the time-step $dt^*=0.0032$ defined in
units of $(a^2m/U_A)^{1/2}$ (that corresponds to $\sim 1.7\cdot
10^{-12} s$ for water-like molecules and to $\sim 2.1\cdot 10^{-12} s$
for argon-like atoms \cite{vilaseca:084507}).
We performed the same check as in Ref. \cite{vilaseca:084507} in order to
verify that the value used for $dt$ is small enough to satisfy the
energy conservation of the system. 
 
The confining parallel walls are placed along the $z$ axis at a
separation distance $L_z$.
The solvophilic wall is composed of a triangular lattice of CSW
particles quenched with the position of the centers placed at
$z_{phil}=0$. The lattice constant is $d=a$. The solvophobic wall has
no structure and is obtained by imposing the repulsive potential
$U_{phob}(z)\equiv(\sigma/z)^9$ where
$z\equiv|z_{particle}-z_{phob}|$, and $z_{phob}=L_z$
\cite{varnik2000}. In the following we consider the parameter
$\sigma=1$.    
We adopt periodic boundary conditions in the $x$ and $y$ directions. 
In order to compute the effective density $\rho_{eff}$, we need to
compute the effective volume $V_{eff}$ accessible to particles
$V_{eff}=L_z^{eff}A$, where $A\equiv L_xL_y$ is the section of the
simulation box and $L_z^{eff}$ is the effective distance between the
plates. By considering the quenched particles forming the philic wall
and the repulsive strength of the potential of the phobic wall, we
obtain $L_z^{eff}\simeq L_z-a/2-(1/T)^{1/9}$ \cite{varnik2000}. 
The effective density is
$\rho^*_{eff}=\rho^*_{eff}(\rho^*,T^*)=\rho^*\cdot(L_z/L_z^{eff})$. 

To explore different densities for the confined system in the $NVT$
ensemble, we change $L_x$ and $L_y$. We keep $L_z$ and $N$ constant to
exclude finite-size effects when we compare results for different
densities. 
For each density, we equilibrate the system by annealing from
$T^*=4$. For each temperature, the system is equilibrated during
$10^6$ time steps. We observe equilibrium after $10^4$ time steps for
the range of $\rho$ and $T$ considered here.
All our results are averaged over $10$ independent samples.

In order to check the stability of the system, we verify that the
energy and the pressure are equilibrated. 
To compute the pressure we follow the approach of
Refs.\cite{Rowlinson1982,varnik2000}.  
Due to the inhomogeneous nature of the system, the pressure is a
tensor ${\bf P}({\bf r})\equiv{\bf P}^K({\bf r})+{\bf P}^U({\bf
  r})+{\bf P}^W({\bf r})$ where ${\bf P}^K({\bf r})\equiv
k_BT\rho({\bf r}){\bf \hat 1}$ is the kinetic contribution as in an
ideal gas, ${\bf \hat 1}$ is the unit tensor, ${\bf P}^U({\bf r})$
is the potential contribution due to the interparticle interaction,
and ${\bf P}^W({\bf r})$ is the contribution of the walls.  
At equilibrium the system is mechanically stable if ${\bf\nabla\cdot
  P}={\bf 0}$. Considering the planar symmetry of the system, the
normal (or orthogonal) and tangential (or parallel) component of the
pressure can be written respectively as: 
$P_{\bot}(z)=P_{zz}=const.$ and  $P_{||}(z)=P_{xx}(z)=P_{yy}(z)$,
while mixed components are zero, i.e., no shear forces are present
\cite{varnik2000}. 

To verify the stability of the system, we computed the
normal component of the pressure as a function of $z$
\begin{equation}\label{equ:pressure_norm}
P_{\bot}(z)\equiv P^K_{zz}(z)+P^U_{zz}(z)+P^W_{zz}(z)
\end{equation} 
The kinetic part is $P^K_{zz}(z)=k_BT\rho(z)$. For the potential part
we used the Todd, Evans and Daivis formulation of the pressure \cite{todd1995}
\begin{equation}\label{equ:pressure_pot}
P^U_{zz}(z)=\frac{1}{2A}\left\langle\sum_{\mbox{${\setstretch{0.6}\begin{array}{c}\scriptstyle
    i=1\\ \scriptstyle (i\neq j)\end{array}}$}}^NF^z_{ij}[\Theta(z_i-z)\Theta(z-z_j) - \Theta(z_j-z)\Theta(z-z_i)]\right\rangle
\end{equation}
where $F^z_{ij}$ is the z-component of the interaction force between
particles $i$ and $j$. The products of Heaviside step functions,
$\Theta$, select couple of particles that lie in different semispace
respect to the plane parallel to the walls with coordinate $z$. 
The walls contribution to the pressure can be computed
from Eq.~\ref{equ:pressure_pot}, with $z_{phil}<z_i<z_{phob}$ for
$i=1,...,N$, as \cite{varnik2000}   
\begin{equation}\label{equ:pressure_walls}
P^W_{zz}(z)=\dfrac{1}{A}\left\langle\sum_{i=1}^{N}\left[F^z_{i,phil}\Theta(z_i-z)-F^z_{phob,i}\Theta(z-z_i)\right]\right\rangle
\end{equation}
where $F^z_{i,phil}\equiv\sum_{j\in phil}F^z_{ij}$ and
$F^z_{phob,i}\equiv-dU_{phob}(z)/dz$ are the interaction forces of the
particle $i$ with the philic wall and the phobic wall, respectively.
\begin{figure}
\begin{center}
\includegraphics[width=16cm]{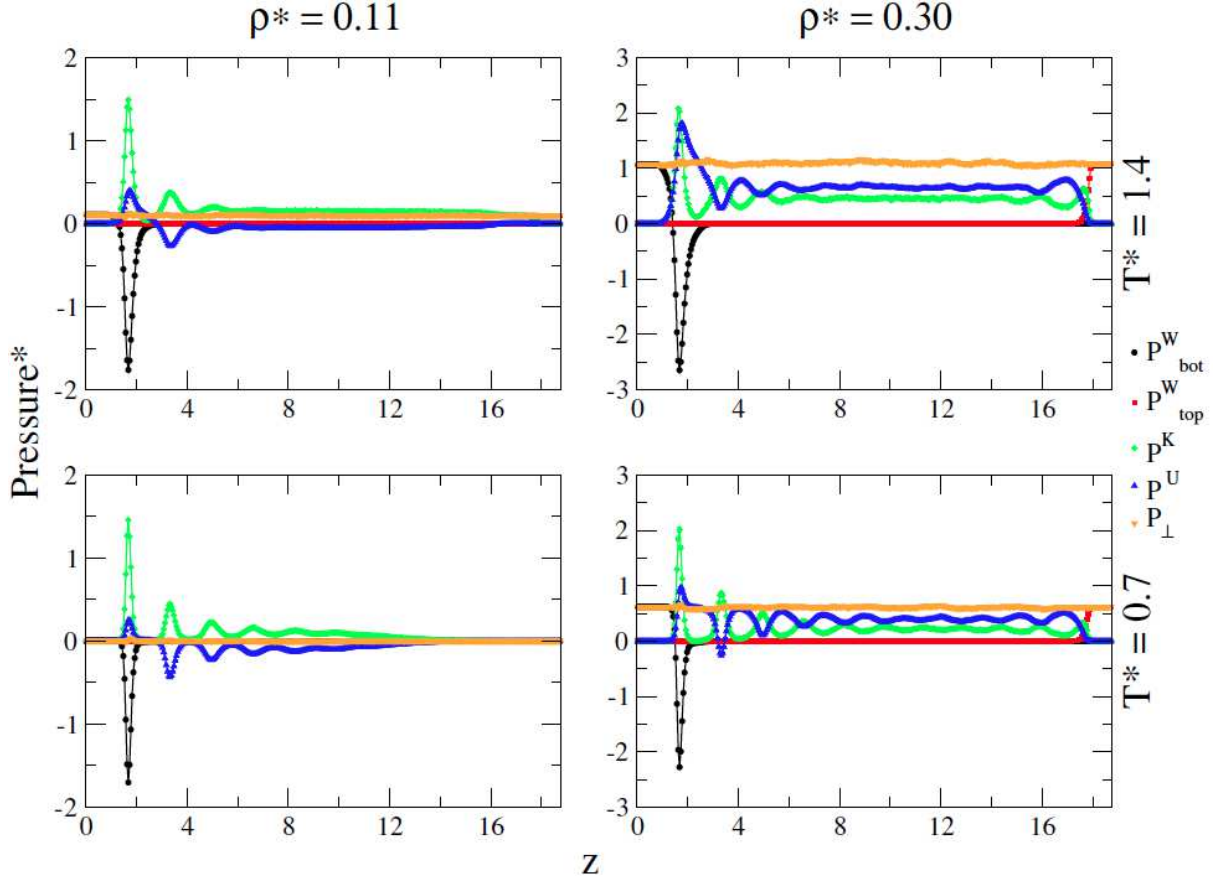}
\vspace{0.3cm}
\caption{\label{fig:pressures} Normal pressure profile $P_{\perp}(z)$,
  for densities $\rho^*=0.11,0.30$ and temperatures
  $T^*=1.4,0.7$. The effective densities are:
  $\rho_{eff}^*(0.11,1.4)=0.1193$, $\rho_{eff}^*(0.30,1.4)=0.3254$,
  $\rho_{eff}^*(0.11,0.7)=0.1198$, $\rho_{eff}^*(0.30,0.7)=0.3268$.  
  Each component of the normal pressure is displayed. In particular,
  ${\bf P}^W_{phil}$ is the contribution of the philic wall, ${\bf
    P}^W_{phob}$ is the contribution of the phobic wall, ${\bf P}^K$
  is the kinetic contribution as in an ideal gas, ${\bf P}^U$ is the
  potential contribution due to the interparticle interaction.}
\end{center}
\end{figure}
The different components compensate each other in order to keep the
normal pressure constant (Fig.\ref{fig:pressures}). 
In particular, the philic wall contributes with a positive term
$P^W_{phil}$ from $z=0$ to $z=R_R/a=1.6$ and with a negative term
from $z=R_R/a=1.6$ to $z=r_c/a=3$, due to the interaction force
between the particles of the fluid and those of the philic wall. 
The phobic wall contributes with a positive term $P^W_{phob}$ from
$z\simeq L_z-(1/T)^{1/9}$ to $z=L_z$ due to the repulsive potential.
The kinetic pressure $P^K(z)$ is a signature of the density profile,
while the potential pressure $P^U(z)$ can contribute with a positive
or negative term depending which part of the interparticle
interaction, repulsive or attractive, respectively, dominate.
After having verified that the system is stable, i.e.,
$P_{\perp}=P_{zz}=const.$, we compute the normal pressure as
$P_{\perp}=P^W_{zz}(z<z_1^*)=\dfrac{1}{A}\left\langle\sum_{i=1}^{N}F^z_{i,phil}\right\rangle=P^W_{zz}(z>z_2^*)=-\dfrac{1}{A}\left\langle\sum_{i=1}^{N}F^z_{i,phob}\right\rangle$  
where $z_1^*$ and $z_2^*$ are z-coordinates sufficiently near the
philic wall and the phobic wall, respectively, for which the kinetic
$P^K_{zz}$ and the potential $P^U_{zz}$ part of the pressure are zero.


\section{Results and discussion}
\label{sec:results}


\subsection{Density profile and aggregation state}
\label{sec:profile}

\begin{sidewaysfigure}
\begin{center}
\vspace{1cm}
\includegraphics[width=18cm]{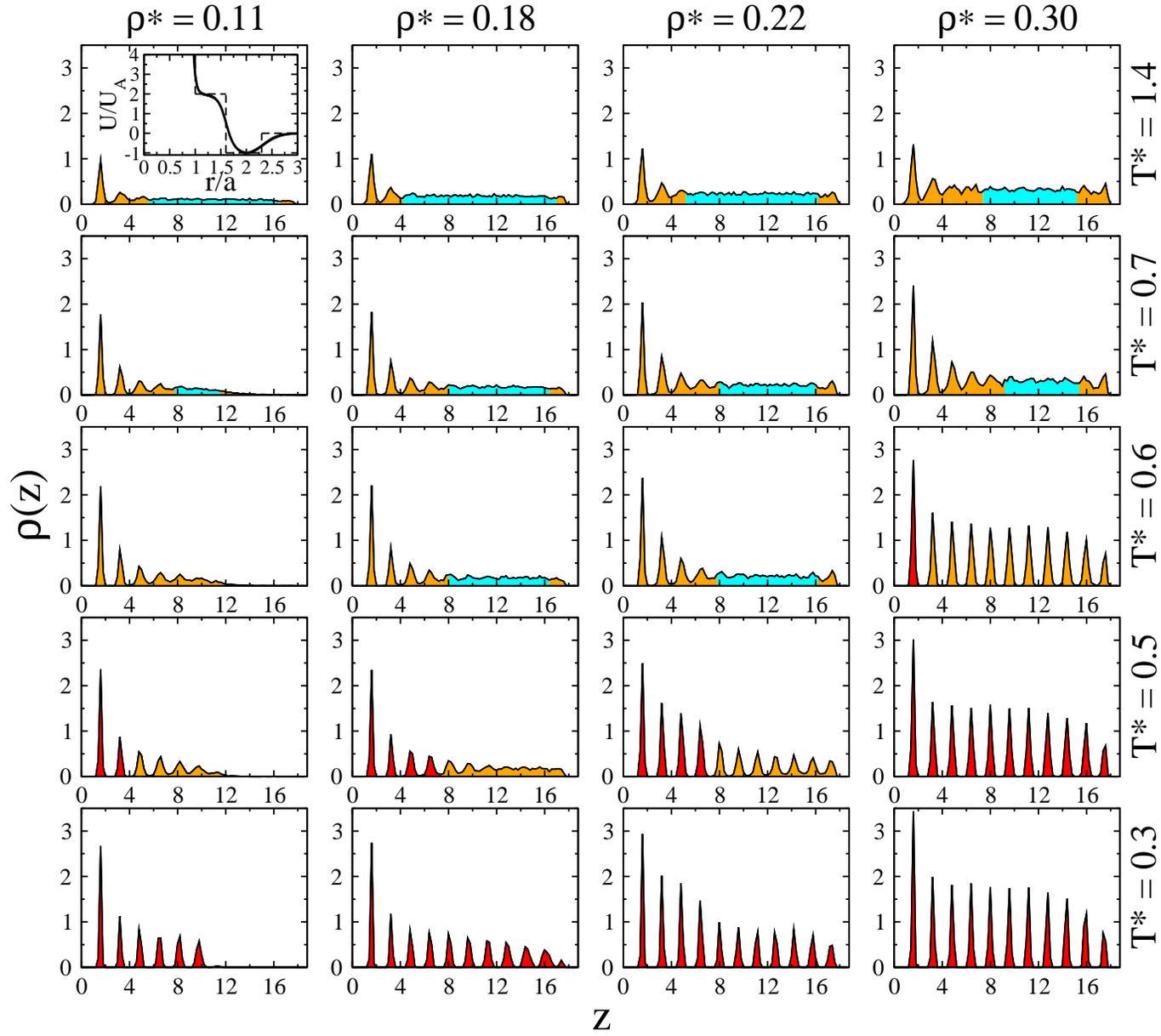}
\end{center}
\caption{\label{fig:density} Density profile $\rho(z)$ for fluid
  densities $\rho^*=0.11, 0.18, 0.22, 0.30$ and temperatures $T^*=1.4,
  0.7, 0.6, 0.5, 0.3$. A binning is performed using a bin equal to
  $0.2$ to get curves composed by $94$ points. We used cyan, orange
  and red color coding to indicate if the system is in a homogeneous
  fluid, heterogeneous fluid and solid phase respectively. Inset: the
  interparticle potential considered in this work.} 
\end{sidewaysfigure}

As pointed out in the introduction (Sec.\ref{sec:introduction}), the
confinement can modify the structure of a fluid resulting in an
inhomogeneous density profile. In a slit pore geometry, near the
confining walls, particles form layers parallel to the walls as the
temperature is decreased or the density is increased, as shown by our 
calculations for the density profile $\rho(z)$
(Fig.~\ref{fig:density}). 

To establish the aggregation state for each layer, we compute, layer
by layer, the lateral radial distribution function $g_{\|}(r_{\||})$;
the 2d Voronoi tessellation of each layer; the mean square
displacement (MSD) and the survival probability (SP) of molecules in
each layer. All these quantities together, as we discusse in the
following, allow us to identify the presence and coesistence of the
solid, heterogeneous fluid and homogeneous fluid. 

Our analysis shows that the system organizes forming layers parallel
to the solvophilic wall at any temperature $T$ and average density
$\rho$. At high $T$ and low $\rho$ we find only one well defined layer
and no layering near the solvophobic wall, while the whole system is in
the fluid state. By decreasing $T$ and increasing $\rho$, the number
of layers increases up to eleven.
Furthermore, at higher value of $\rho$, the layers appear also near
the solvophobic wall. However, for high enough $T$ and $\rho$ we
observe that away from the walls the system is in the fluid
state. Nonetheless, the walls affect the density profile, changing
system density and aggregation state, over an extension that in our
case can be up to $11$ layers that depends on temperature and average
density of the system.


\subsection{Radial distribution function and spatial configuration analysis}
\label{sec:gr}

The in-layer radial distribution function $g^n_{\|}(r_{\|})$ for the
$n$-th layer (with $n=1,2,...,11$) is computed as
\begin{equation}
g^n_{\|}(r_{\|})\equiv\dfrac{1}{(\rho^n)^2A\delta z}\sum_{i\ne
  j}\delta(r_{\|}-(r_{ij})_{\|})\left[\Theta\left(\dfrac{\delta z}{2} -|z_i-z^n|\right)\Theta\left(\dfrac{\delta z}{2} -|z_j-z^n|\right)\right]
\end{equation}
where $\rho^n$ and $z^n$ are the density of particles and the
z-coordinate of the layer $n$, respectively, $r_{||}=(x^2+y^2)^{1/2}$
is the transverse distance between two particles in the same layer
(and in internal units is $r^*_{\|}=r_{\|}/a$).  
The Heaviside step functions, $\Theta$, select couple of particles that
lie in the layer $n$ of width $\delta z$. 
The $g^n_{\|}(r_{\|})$ is proportional to the probability of finding a
molecule in the layer $n$ at a distance 
$r_{\|}$ from a randomly chosen molecule of the same layer $n$. 
The definition of the layer $n$ in which lies a
particle $i$ is once for all established according to the value of
the z-coordinate of the particle $i$ as: for the first layer 
$0<z_i^{n=1}<(3/2)\delta_z$, with $\delta_z=R_R/a$, and
$(j-1/2)\delta_z<z_i^{n=j>1}<(j+1/2)\delta_z$ for the others.
This is a natural choice because particles at low temperatures or high
densities tend to stratify in layers whose interdistance is
approximately equal to $\delta_z$. 

\begin{sidewaysfigure}
\begin{center}
\begin{minipage}{22cm}
\centering
\raisebox{-1\height}{\includegraphics[height=14.5cm]{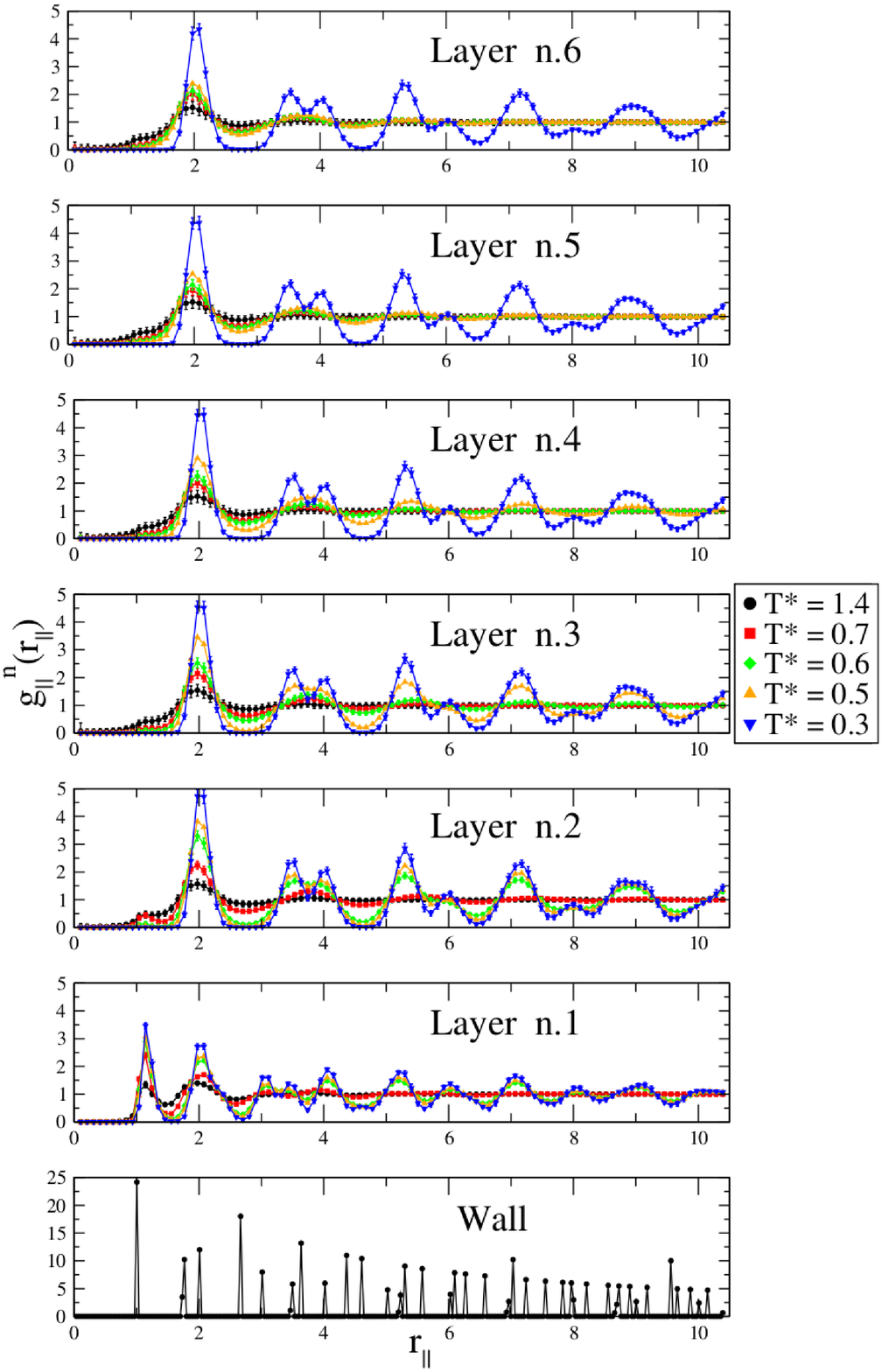}}
\hspace*{0.05cm}
\raisebox{-0.97\height}{\includegraphics[height=14.5cm]{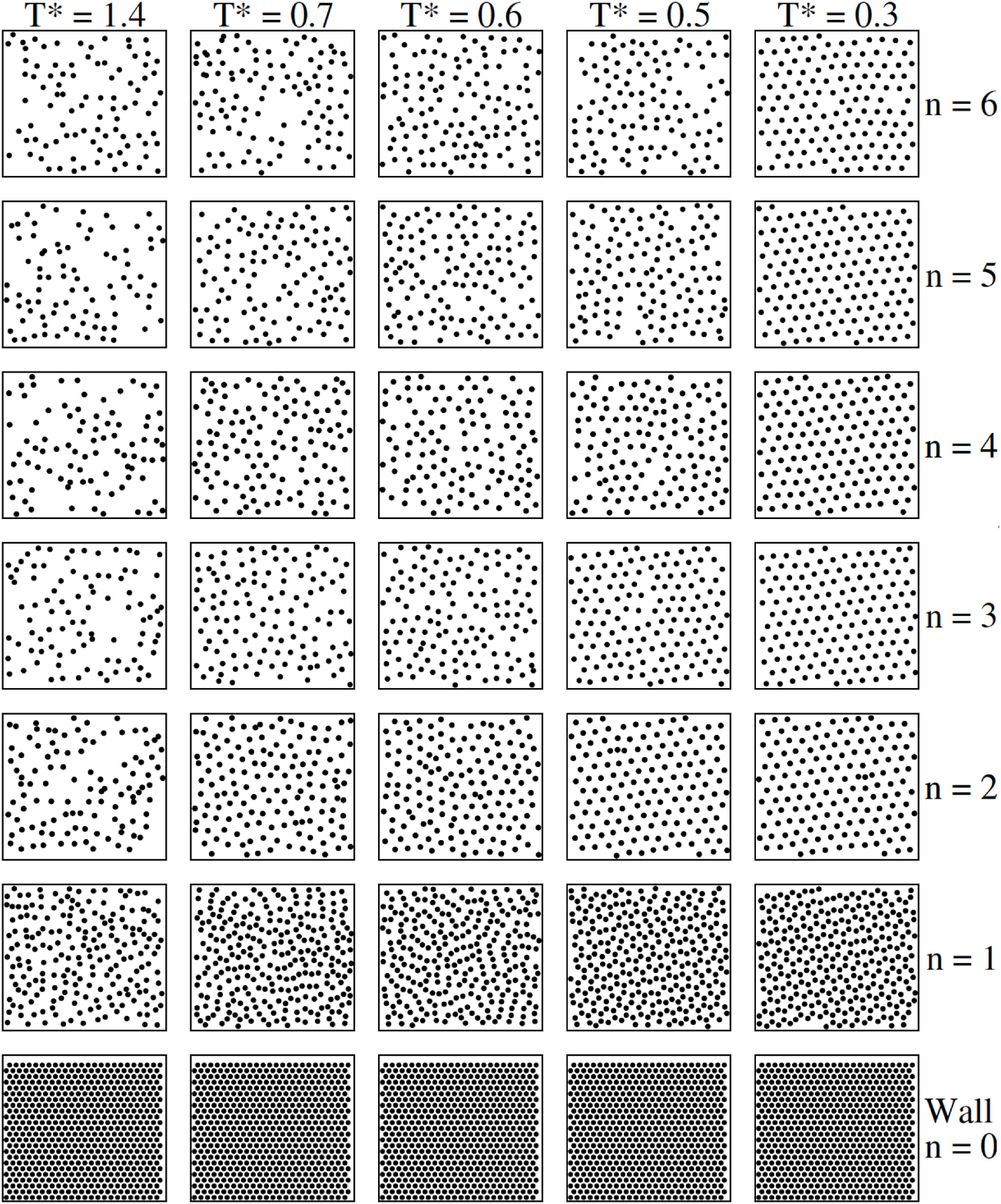}}
\end{minipage}
\end{center}
\vspace{0.3cm}
\caption{\label{fig:gr_rho011} Left-most panels: in-layer radial
  distribution function $g_\|^n(r_\|)$ for $\rho^*=0.11$ for the
  particles in the solvophilic wall and the first six layers (from
  bottom to top panel).  
  All the other panels show tipical particle configurations after
  $t=10^6$ simulation steps for $T^*$ as indicated by the top most
  labels and for the layers indicated by the right most
  label. The effective densities (see Sec.\ref{sec:simulations}) that
  correspond to $\rho^*=0.11$ and temperatures
  $T^*=1.4,0.7,0.6,0.5,0.3$ are 
  $\rho^*_{eff}=0.1193,0.1198,0.1200,0.1201,0.1205$, respectively.} 
\end{sidewaysfigure}

At low density ($\rho^*=0.11$, Fig.\ref{fig:gr_rho011}), we observe
that the system is in a fluid state for high temperature ($T^*=1.4$)
in any layer. The
$g^n_{\|}(r_{\|})$ of the layer $n=1$ shows a first peak around the
shoulder radius ($r^*_{\|}\simeq 1.6$) and a second peak around the
attractive well radius ($r^*_{\|}\simeq 2$), while for the other layers
only the second peak is present. 
We interpret this difference between the first and the other layers as
the consequence of a ``templating'' effect of the solvophilic wall
that at high $T^*$ is observed only on the first layer.   

At $T^*=0.7$ the system shows the same behavior observed for higher
temperatures, except that the layer near the solvophilic wall develops
patterns. This behavior is reminiscent of what has been observed in
monolayers with an interparticle potential composed by a hard core
and a soft repulsive shoulder \cite{MalPel2003}. 

At $T^*=0.6$ the layer near the solvophilic wall is still showing
patterns, while the layer $n=2$ is forming crystal patches. 
This is evident from the analysis of the $g_{\|}^{n=2}$ that goes to
zero for $r^*_{\|}\simeq 2.75$ and $4.75$ at $T^*=0.6$, consistent with
an incipient triangular crystal with lattice step given by the
interaction potential attractive distance $R^*_A=2$.
The other layers are in a fluid state. 

At $T^*=0.5$ the first layer is forming a hexagonal crystal.
Although the hexagonal crystal in $n=1$ does not overlap exactly with
the triangular wall structure, the comparison of the $g_{\|}^{n=1}$
and $g_{\|}^{n=0}$ of the wall shows a strong correlation between the
two structures, suggesting a ``templating'' effect.
This effect due to the attraction to the solvophilic wall is so strong
at $T^*=0.5$ that is forcing particles to be at their repulsive
distance. 
The high-energy cost of the resulting honeycomb lattice forming in the
layer $n=1$ is compensated by the large number of attractive
interactions between the particles at $n=1$ and those of the wall
($n=0$) from one hand, and between the particles themeselves at $n=1$
from the other hand.
This free energy minimization process is analyzed in
Sec.\ref{sec:structural} in the discreet potential approximation to
understand the stripe phase formation.
The triangular structure that was incipient for $n=2$ at
high $T^*$, for $T^*=0.5$ is well defined for $n=2$ and $n=3$, with
defects in the layer $n=3$. This triangular structure is the dual
lattice of the $n=1$ hexagonal layer and its formation is the
consequence of a ``molding'' effect of the layer $n=1$ onto the layer
$n=2$.   
Note that while the wall ($n=0$) layer has a templating effect on the
$n=1$ layer, the $n=1$ layer has a molding effect on the $n=2$
layer. The difference between the two cases is due to the smaller
density of the $n=1$ layer with respect to that of the wall. The
smaller density does not allow to compensate the high energy cost of
the propagation of the hexagonal crystal to the layer $n=2$. On the
other hand, the triangular crystal of the $n=2$ layer is energetically
favorable, because the particles are all at the attractive distance,
and at this temperature can propagate to the $n=3$ layer again with a
``templating'' effect. 
The layer $n=4$ is made of a few triangular crystallites immersed in
the fluid, while the other layers are in a fluid state.
  
At $T^*=0.3$ both the templating and the molding effect are
stronger. In particular the template of the $n=2$ layer propagates
over all the six layers that are formed at this density and
temperature.  

\begin{sidewaysfigure}
\begin{center}
\begin{minipage}{22cm}
\centering
\raisebox{-1\height}{\includegraphics[height=14.75cm]{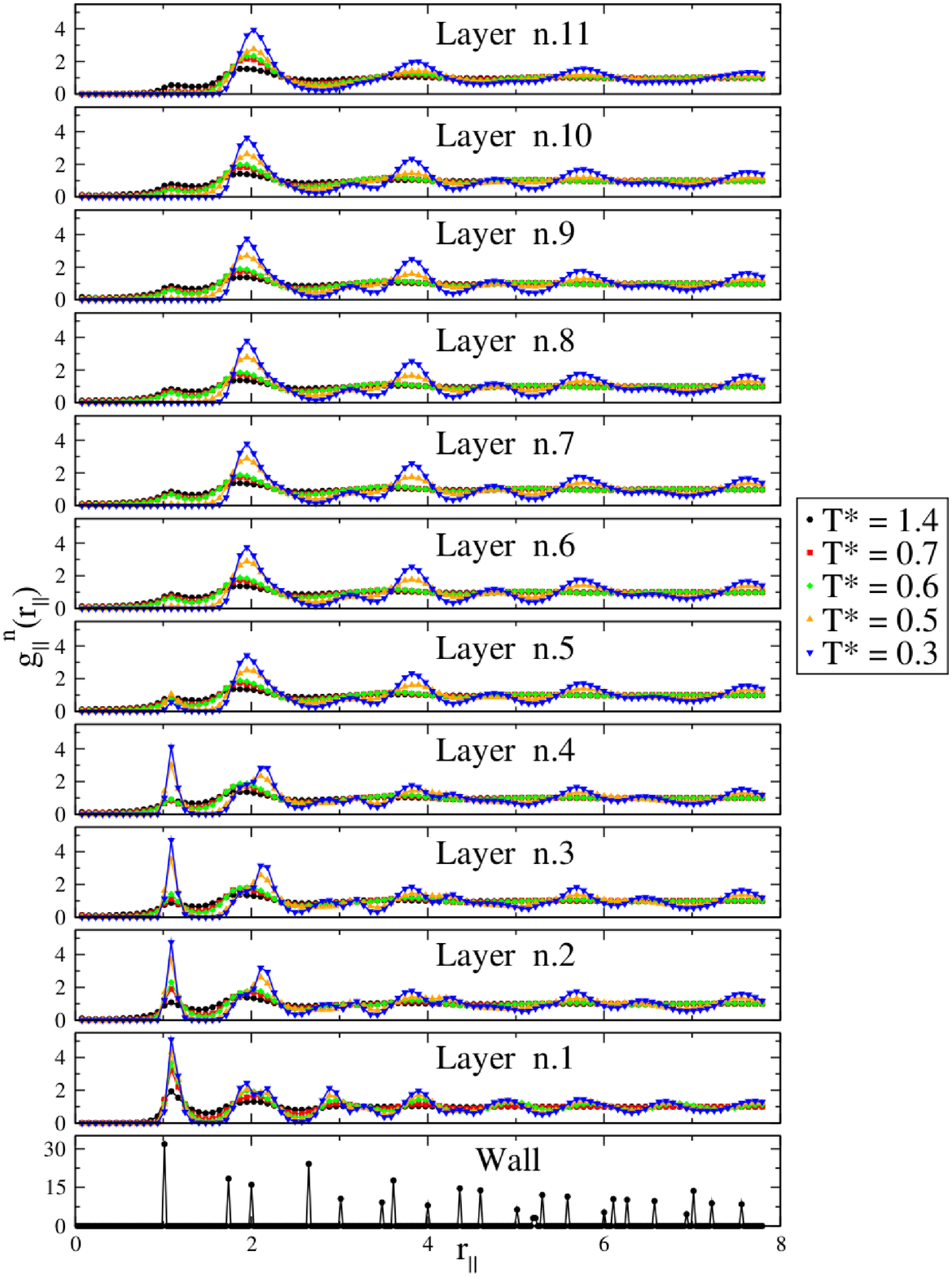}}
\hspace*{0.15cm}
\raisebox{-0.975\height}{\includegraphics[height=14.75cm]{config_layers_rho022_v4.eps}}
\end{minipage}
\end{center}
\vspace{0.3cm}
\caption{\label{fig:gr_rho022} As Fig.\ref{fig:gr_rho011}, but for
  $\rho^*=0.22$. The effective densities (see
  Sec.\ref{sec:simulations}) that correspond to $\rho^*=0.22$ and
  temperatures $T^*=1.4,0.7,0.6,0.5,0.3$ are 
  $\rho^*_{eff}=0.2386,0.2397,0.2399,0.2402,0.2411$, respectively.}
\end{sidewaysfigure}

At intermediate density ($\rho^*=0.22$, Fig.\ref{fig:gr_rho022}), for
$T^*=1.4$ and $T^*=0.7$ we observe the same qualitative behavior as
for the low density case.
For $T^*=0.6$ the layer $n=1$ has less tendency to form patterns
respect to the low density case, and the layer $n=2$ to order in a
crystal structure. Therefore, the confined system is more fluid at
this density than at lower density. We understand this result as a
consequence of the larger hydration at higer density.
   
At $T^*=0.5$ the first layer has partially crystallized in the
hexagonal and partially in the triangular structure following the
template of the wall. Therefore, the templating effect is now stronger
then the corresponding case at lower density. The hexagonal crystal
shows now a preferred direction of symmetry. This direction propagates
to the layer $n=2$, where we observe stripes along the preferred
direction. The stripes propagate up to $n=4$ layer, while the other
layers are in a fluid state. 
The peak of $g^n_{\|}(r_{\|})$ at $r^*_{\|}\simeq 2.1$ (that corresponds
to the average second nearest neighbor distance) is a signature of the
stripe phase formation.  

At $T^*=0.3$ the preferred direction in the deformation of the hexagonal
crystal for $n=1$ is more evident and we observe a clear stripe phase
for the layers from $n=2$ to $n=4$, with a peak of $g^n_{\|}(r_{\|})$
at $r^*_{\|}\simeq 2.1$ more pronounced than the case at $T^*=0.5$.
The other layers form a triangular crystal at the attractive distance.   

\begin{sidewaysfigure}
\begin{center}
\begin{minipage}{22cm}
\centering
\raisebox{-1\height}{\includegraphics[height=15.5cm]{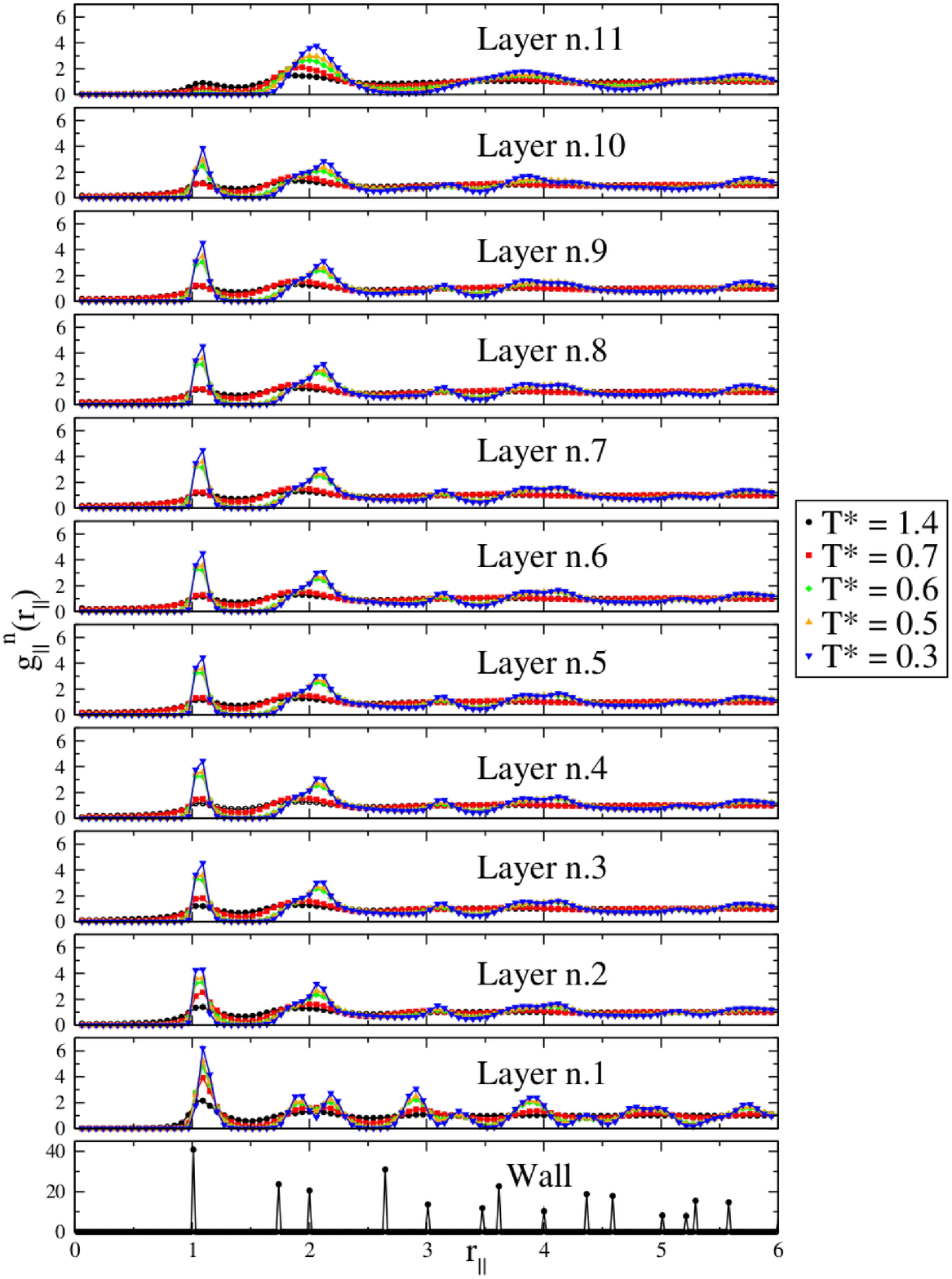}}
\hspace*{0.15cm}
\raisebox{-0.965\height}{\includegraphics[height=15.5cm]{config_layers_rho030_v4.eps}}
\end{minipage}
\end{center}
\vspace{0.3cm}
\caption{\label{fig:gr_rho030}  As Fig.\ref{fig:gr_rho011}, but for
  $\rho^*=0.30$. The effective densities (see
  Sec.\ref{sec:simulations}) that correspond to $\rho^*=0.30$ and
  temperatures $T^*=1.4,0.7,0.6,0.5,0.3$ are
  $\rho^*_{eff}=0.3254,0.3268,0.3272,0.3276,0.3288$, respectively.}
\end{sidewaysfigure}

At high density ($\rho^*=0.30$, Fig.\ref{fig:gr_rho030}), for
$T^*=1.4$ we observe the same qualitative behavior as for the lower
density cases.
At $T^*=0.7$ we found that the only difference with the lower density
case is that the layer $n=1$ is forming crystallites following the
template of the wall.
At $T^*=0.6$ the layer $n=1$ has a different and incipient crystal
structure (Kagome lattice) with defects that is better defined at
lower $T$. This is evident from the analysis of the $g_{\|}^{n=1}$ that 
goes to zero for $r^*_{\|}\simeq 1.6$ and $2.75$ at $T^*=0.6$. 
The layer $n=2$ shows patterns very close to the stripe configuration.
The corresponding $g_{\|}^{n=2}$ goes to zero for $r^*_{\|}\simeq 1.5$
and shows a peak for $r^*_{\|}\simeq 2.1$. These characteristics of the
$g_{\|}^{n}$ are typical of a stripe phase.
The layer $n=3$ and $n=4$ still show patterns close to the stripe
phase, but in a less pronounced way. 
From the layer $n=5$ to the $n=10$ the pattern is vanishing.
The layer $n=11$ is showing an incipient triangular crystal with
lattice step given by the interaction potential attractive distance
$R^*_A=2$. This is evident from the analysis of the $g_{\|}^{n=11}$
that approaches zero for $r^*_{\|}\simeq 2.75$ and $4.75$ at $T^*=0.6$.

At $T^*=0.5$ the layer $n=1$ is forming a kagome crystal with defects.
The layers from $n=2$ to $n=10$ show a stripe phase and the layer
$n=11$ is forming a triangular crystal with defects.

At $T^*=0.3$ the Kagome crystal of layer $n=1$ has no defects.
The layers from $n=2$ to $n=10$ are in a stripe phase and the layer
$n=11$ is forming a well defined triangular crystal.

\begin{figure}[H]
\begin{center}
\begin{minipage}{13cm}
\centering
\hspace*{-0.5cm}\raisebox{-0.038\height}{\includegraphics[height=18.5cm]{gr_first_layer_all_densitiesD.eps}}
\hspace*{0.5cm}
\raisebox{0\height}{\includegraphics[height=18cm,width=3.7cm]{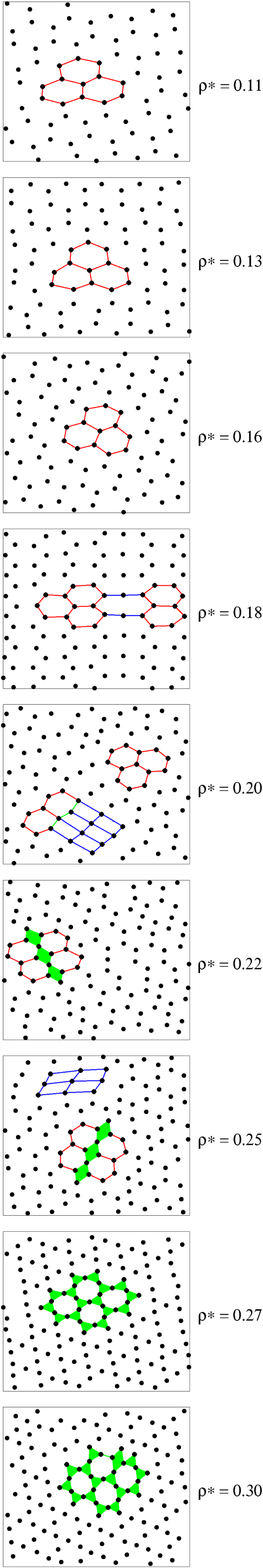}}
\end{minipage}
\end{center}
\caption{\label{fig:layers_gr_snap} Radial distribution function $g^{n=1}_\|
  (r_\|)$ (left) and snapshots (right) computed for the first fluid layer
  for temperature $T^*=0.3$ and densities (from top to bottom)
  $\rho^*=0.11, 0.13, 0.16, 0.18, 0.20, 0.22, 0.25, 0.27, 0.30$ that
  correspond to the effective densities
  $\rho^*_{eff}=0.1205,0.1425,0.1753,0.1973,0.2192,0.2411,0.2740,0.2959,0.3288$,
  respectively.}
\end{figure}

As discussed above, the layer $n=1$ close to the solvophilic wall
($n=0$) is subjects to the templating effect for all densities at low
temperatures. 
In Fig.\ref{fig:layers_gr_snap} we compare the $g^{n=1}_{\|}(r_{\|})$ and
the snapshots of the first layer for $T^*=0.3$ at several densities.
In order to compare layers, that correspond to different densities,
between them, we considered a portion of each layer of the same size
($L_x$x$L_y$ of the system at $\rho^*=0.30$).     
For densities $\rho^*=0.11, 0.13, 0.16$ the first layer is forming a
distorted hexagonal lattice characterized by a $g^{n=1}_{\|}(r_{\|})$
with a first peak at $r^*_{\|}/a\sim 1.15$ and vanishing for
$r^*_{\|}\simeq 1.6$ and $2.75$. 
By increasing $\rho$ we observe a progressive shift to higher values
of $r_{\|}$ of all the peaks of $g^{n=1}_{\|}(r_{\|})$, but the first
that, instead, is becoming more pronounced as a consequence of a
better local order.
For densities between $\rho^*=0.20$ and $\rho^*=0.25$ the first layer
shows a polycrystal phase with coexistence of triangular and square
lattices.    
This corresponds to an intermediate stage toward the well defined
Kagome lattice that is formed for $\rho^*\geq 0.27$, as showed by the
splitting of the second peak of $g^{n=1}_{\|}(r_{\|})$ into two close
peaks at $r^*_{\|}\simeq 1.9$ and $2.2$.
%
%
%
%


\subsection{Mean square displacement analysis}
\label{sec:MSD}

In order to characterize space-dependent diffusion properties of our
system, we compute the mean square displacement (MSD) associated to
each layer of the slit. 
We observe that, except for low temperatures, a particle can visit
different layers in which the aggregation state can change from
homogeneous to heterogeneous liquid and vice versa.
For this reason we calculate the MSD only for those particles that
remain in a layer over the entire time interval under consideration
and we average over all possible time interval. 
Therefore, the MSD associated to each layer $n$ is defined as 
\begin{equation}\label{equ:MSD}
\langle (\Delta r^n_{||}(\tau))^2\rangle\simeq\langle
(r^n_{||}(t-t_0)-r^n_{||}(t_0))^2\rangle  
\end{equation}
where $\tau\simeq t-t_0$ is the time spent in the layer $n$ by a
particle that entered in the layer at time $t_0$. 
\begin{sidewaysfigure}
\begin{center}
\hspace*{0cm}\includegraphics[width=19cm]{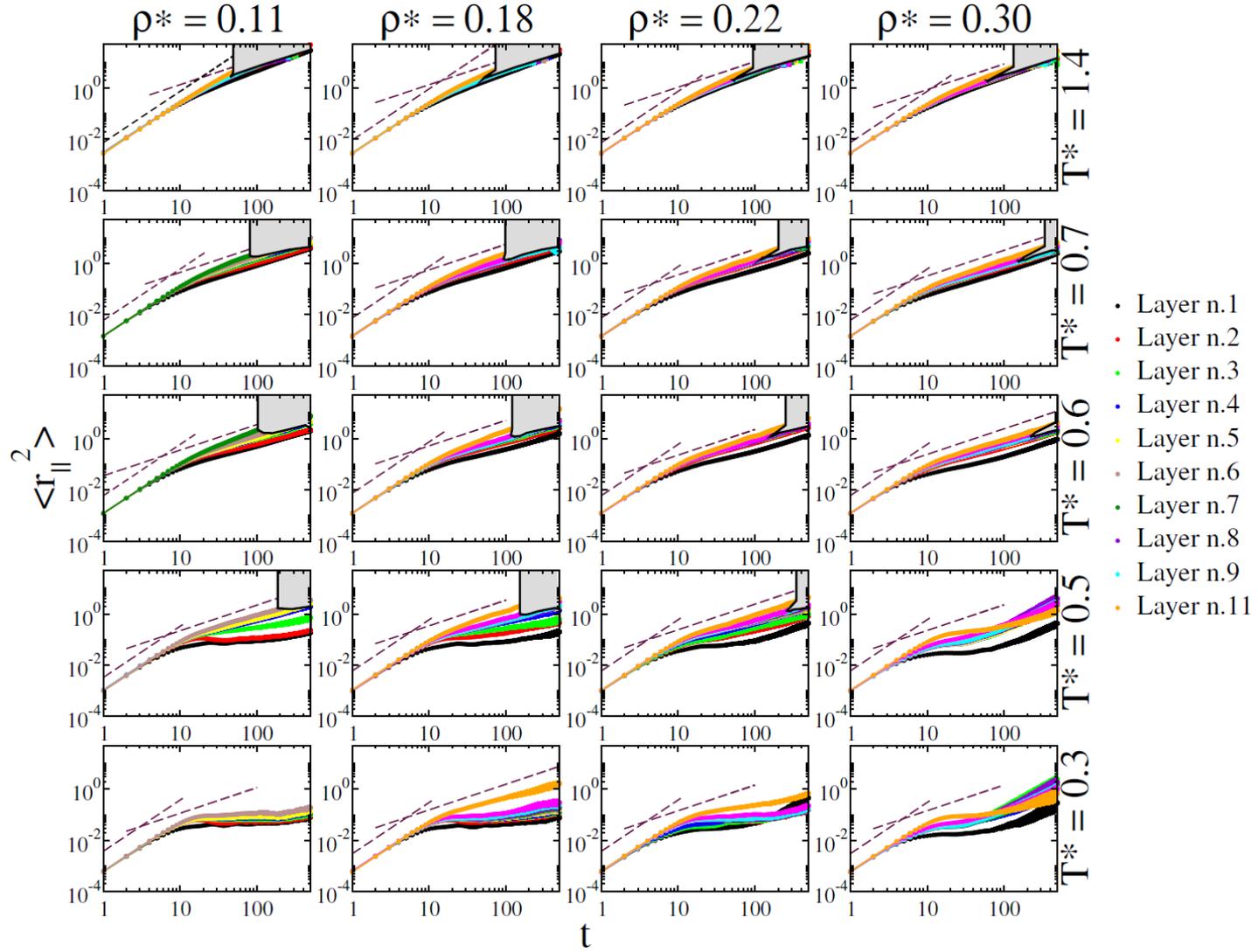}
\end{center}
\vspace{0.3cm}
\caption{\label{fig:diff_layers} In-plane mean square displacement
  (MSD), $\langle r_{||}^2\rangle$, as a function of time, $t$, for
  fluid densities $\rho^*=0.11, 0.18, 0.22, 0.30$ and temperatures
  $T^*=1.4, 0.7, 0.6, 0.5, 0.3$. Different colors represent different
  layers, as indicated in the legend. Dashed lines represent the
  ballistic and diffusive regimes at early and long times,
  respectively. The gray regions give an indication of the time
  interval over which the MSD is not well defined, as discussed in
  Sec.\ref{sec:SP}.}
\end{sidewaysfigure}
In according to the standard definition of the MSD,
$\lim_{\tau\rightarrow\infty}\langle(\Delta
r_{\|}^n(\tau))^2\rangle=4D_{\|}\tau^{\alpha}$ where $D_{\|}$ is the
lateral, or parallel, diffusion coefficient and $\alpha$ the diffusion
exponent. The value $\alpha=0$ means that the system is arrested, as
in a solid state where particles can only vibrate around theirs
equilibrium positions;  
for $0<\alpha<1$ the system is subdiffusive corresponding in general
to particles diffusing in complex structures (with non trivial 
microscopic disorder); $\alpha=1$ is the standard diffusive behavior
as in a normal fluid state. For $\alpha>1$ the system is superdiffusive. 
On the other hand, for early times free diffusion we expect the
ballistic regime with $\alpha=2$.
Our analysis (Fig.\ref{fig:diff_layers}) shows that the in-layer MSD
has always a ballistic regime for $t^*\leq 10$.
The corresponding mean displacement is approximately half particle
diameter at high $T$ and low $\rho$ and weakly decreases for increasing
$\rho$ and decreasing $T$ corresponding to the expected decrease of
the mean free path of the particles.

For $T^*\geq 1.4$ all the layers reach the diffusive ($\alpha=1$)
behavior for long times.
By decreasing the temperature the behavior of the layers becomes more
heterogeneous. 
In particular, we observe that the layer $n=1$ near to the solvophilic
wall slows down in a sensible way with respect to the layers at
$T^*\leq 0.7$ and becomes arrested for $T^*\leq 0.5$. At these
temperatures the other layers, including the one near the solvophobic
wall, are diffusive at low densities.   
However, at $\rho^*=0.30$ and $T^*=0.5$ all the layers develop the
plateau in the MSD typical of glassy dynamics.
This behavior is reminiscent of the caging effect in glasses
where the plateau in the MSD is followed by a diffusive regime.       
Here, instead, at $T^*=0.5$ and $\rho^*=0.30$ we observe that for all
the layers but the one near the solvophobic wall ($n=11$), after the
plateau, the dynamics enters in a superdiffusive regime with
$1<\alpha<2$.  
This effect is related to the presence of defects and of a nonuniform
stress field, as discussed in Sec.\ref{sec:structural}. 
For $T^*=0.3$ we observe that all the layers are arrested at
$\rho^*=0.11$. At this low density the slit is only partially filled
(Fig.\ref{fig:diff_layers}). At $T^*=0.3$ and $\rho^*=0.18$ and
$\rho^*=0.22$ also the layer $n=11$ near the solvophobic wall is
present and it is characterized by a larger MSD with respect to the
other layers and by a diffusive regime at long times. The other layers 
have an arrested dynamics.
At $T^*=0.3$ and $\rho^*=0.30$ all the layers from $n=1$ to $n=10$
have a superdiffusive regime at long times, while the layer $n=11$
reaches the diffusive log-time regime. However, its MSD is smaller
than that of the other layer for very long times ($t^*\geq 100$).


\subsection{Survival probability function analysis}
\label{sec:SP}

\begin{sidewaysfigure}
\begin{center}
\hspace*{0cm}\includegraphics[width=19cm]{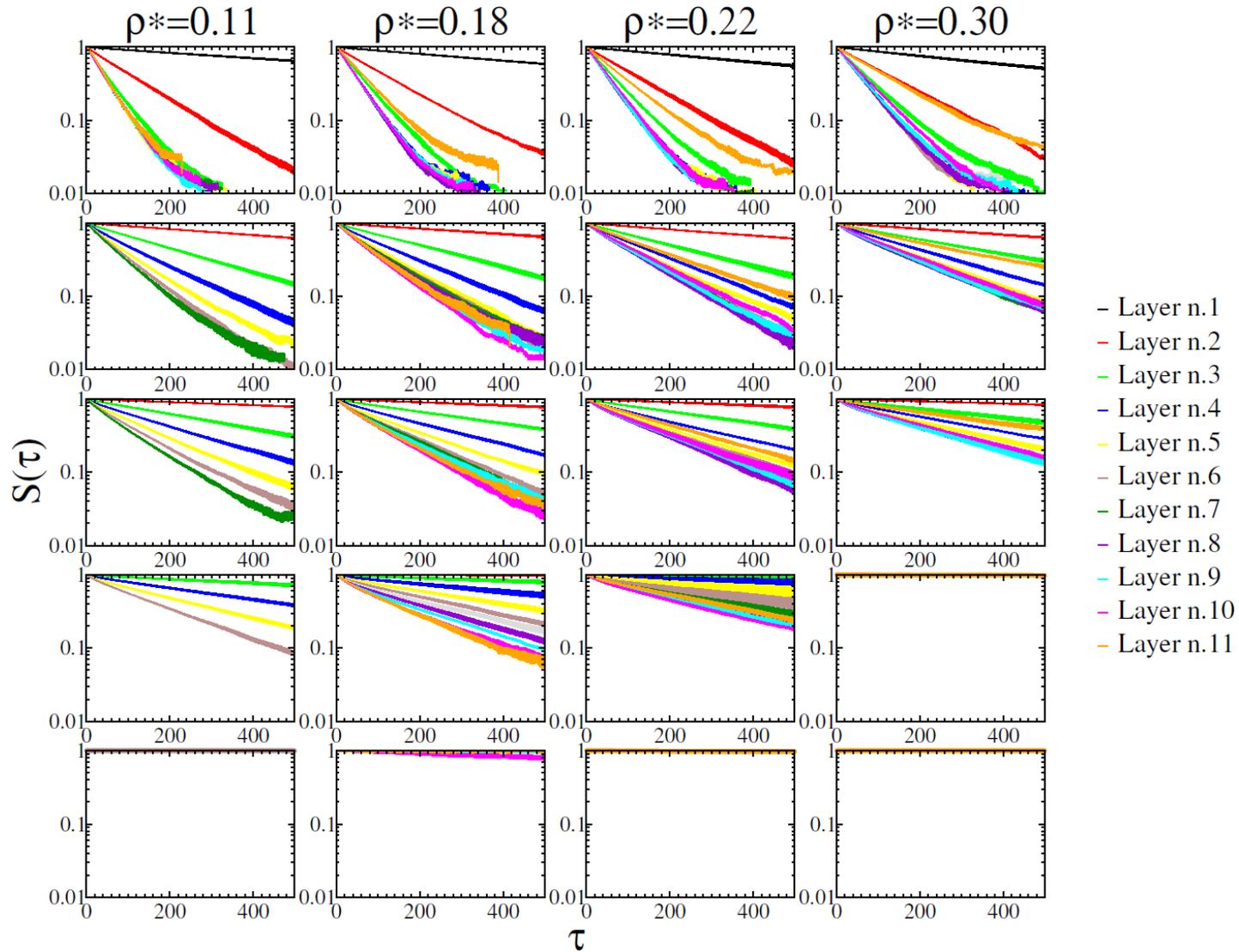}
\end{center}
\vspace{0.3cm}
\caption{\label{fig:Stau} The in-plane survival probability (SP),
  $S(\tau)$, as a function of the time interval $\tau$,
  for fluid densities $\rho^*=0.11, 0.18, 0.22, 0.30$ and temperatures
  $T^*=1.4, 0.7, 0.6, 0.5, 0.3$. Lines with different colors
  correspond to different layers, as indicated in the legend. The
  layers with a faster decay of $S(\tau)$ are those away from the two
  walls of the slit.}
\end{sidewaysfigure}

\begin{figure}[H]
\begin{center}
\includegraphics[width=10cm]{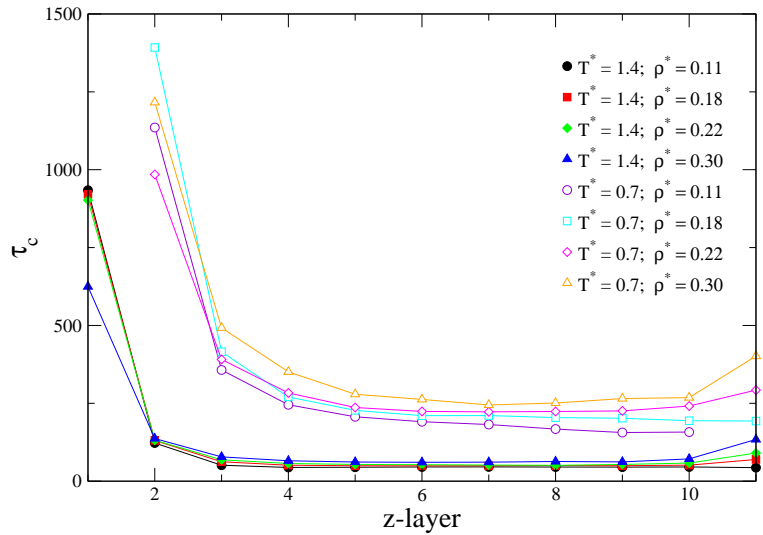}
\end{center}
\vspace{0.3cm}
\caption{\label{fig:tau_layers} Characteristic time decay $\tau_{max}$
  as a function of fluid layer for density
  $\rho^*=0.11,0.18,0.22,0.30$ and temperature $T^*=1.4,0.7$. For
  clarity, the points for $T^*=0.7$ are shifted up by 100 units.}
\end{figure}

As the time proceeds, the average in Eq.~\ref{equ:MSD} for the MSD is
performed on a decreasing number of particles because some of them can
leave the layer. {\it A priori} this reduction of the statistics is not
homogeneous, that means that in general there can be a correlation
between particles that leave the layer and theirs properties, as theirs
velocity components.
Therefore, for $T^*\geq 0.5$, low enough $\rho$, and for the most
diffusive layers there is a time, $\tau_{max}$, after which the
in-layer MSD (Fig.\ref{fig:diff_layers}) is not well defined. 
To estimate $\tau_{max}$ as function of $T$ and $\rho$ for different
layers we analyse the population relaxation of particles in each
layer. In particular, we compute the survival probability (SP)
function, $S^i(\tau)$, which is the probability that a given particle  
stay in the layer $i$ for a time interval $\tau$.
The SP can be calculated as
\begin{equation}\label{equ:SP}
S^i(\tau)\equiv\left\langle\dfrac{N^i(t,t+\tau)}{N^i(t)}\right\rangle 
\end{equation}
where $N^i(t)$ is the number of particle in the layer $i$ at time $t$
and $N^i(t,t+\tau)$ is the number of particle that do not leave the
layer $i$ during the time interval $[t,t+\tau]$.
The SP give an indication of the time interval $\tau_{max}$ over which
the MSD is well defined. 
We observe that $S(\tau)$ has an exponential decay in our
simulations (Fig.\ref{fig:Stau}). We, therefore, define $\tau_{max}$ as
the characteristic decay time $S(\tau)\sim e^{-\tau/\tau_{max}}$.
This choice is consistent with the observation that the MSD in
Fig.\ref{fig:diff_layers} is well defined when $S(\tau)\geq
1/\mbox{e}$. 
We observe that for $T^*\geq 0.6$ and all the densities and for
$T^*=0.5$ and $\rho^*\leq 0.22$, the SP decay is slower for the layer
$n=1$ near the solvophilic wall and becomes faster for the layers away
from the two walls. When the layer $n=11$ near the solvophobic wall is
present, we observe that it has a decay in SP slower than those layers
that are farther away from the wall. At $T^*=0.3$ for all densities,
and at $T^*=0.5$ for $\rho^*=0.30$, there is no decay in SP, consistent
with the crystallization of the layers. 
The non monotonic behavior of $\tau_{max}$ is reported in
Fig.\ref{fig:tau_layers} as a function of layers for different densities
and temperatures.
By comparing Fig.\ref{fig:tau_layers} and Fig.\ref{fig:density} we
observe that $\tau_{max}$ increases when the layers are more
structured in the $z$ direction. Hence, both walls facilitate the
stratification of the fluid, although the structureless phobic wall
does it in a less strong way with respect to the structured
solvophilic wall. 

\vspace{1cm}


\subsection{Liquid veins}
\label{sec:veins}

In Sec.\ref{sec:MSD}, analysing the MSD layer by layer, we have seen
(Fig.\ref{fig:diff_layers}) that for high densities and low
temperatures, after a plateau, the dynamics can enter in a
superdiffusive regime with $1<\alpha<2$.
In this section we show how this behavior is due to the formation of
liquid ``veins'' in such layers.  

The formation of liquid veins is of particular interest in ice during
the freezing of water. Recently, experiments and simulations showed the
presence of liquid water between nanometer-sized ice crystal
\cite{banerjee2013}. 
In polycrystalline systems, the liquid is found along intergranular
junctions, as grain boundaries (see \cite{banerjee2013} and references
therein for the case of water).
Residual stress in these polycrystal structures can be localized along
integranular junctions, and can results in an effective force that
acts on fluid particles present in these junctions. 
The origin of the residual stress in our system is due to the fact
that when the fluid solidifies as the temperature is decreased, the
minimization process of the free energy take place locally, instead of 
globally. In glass forming liquids, this effect is caused by a fast
cooling, while in our system it is due to the layering of the fluid
caused by the confinement. 

\begin{figure}[H]
\begin{center}
\includegraphics[width=11cm]{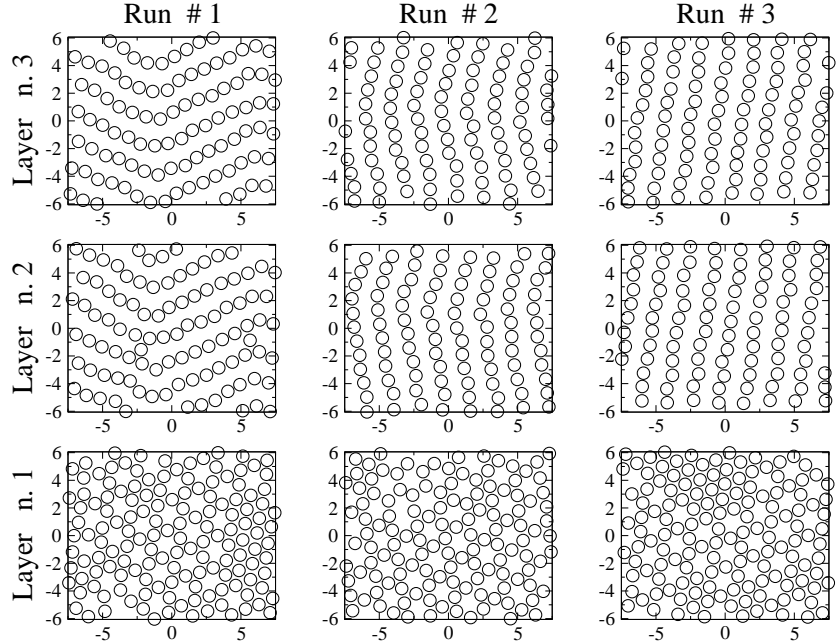}
\end{center}
\vspace{0.3cm}
\caption{\label{fig:snap_z1-2-3_rand1-2-3} Spatial configuration of
  the first three layers of the system close to the solvophilic wall,
  for three different runs (i.e. for three different 
  realization of the initial conditions) at $T^*=0.3$ and
  $\rho^*=0.30$. The size of circles, representing particle positions,
  are chosen to be equal to the particles hard core diameter $a$.}
\end{figure}

In Fig.\ref{fig:snap_z1-2-3_rand1-2-3} we show the spatial
configuration of the first three layers of the system close to the
philic wall, for three different runs (i.e. for three different
realization of initial conditions) at $T^*=0.3$ and $\rho^*=0.30$. 
We observe that particles in the first layer ($n=1$) are characterized
by the same MSD, while particles in other layers ($n=2,3$) can have
different MSD.
In particular in some configurations, we observe veins with
mobility higher than the rest of the system
(Fig.\ref{fig:config_t_z2-z3}). 
We analyzed the trajectories of the particles of these specific
realizations of the system (Fig.\ref{fig:config_t_z2-z3}). 
For these cases we find that these particles with a MSD higher than
the majority belong to the same stripe and diffuse along the stripe
itself. 
We observe that the majority of particles in the layer $n=2$ (and in a
less evident way for the layer $n=3$ (Fig.\ref{fig:config_t_z2-z3}b),
remain spatially localized during the entire simulation, except those
belonging to two stripes moving in the same direction as along stripe
veins (Fig.\ref{fig:config_t_z2-z3}a). 
We observe a similar situation for the layer $n=3$, but here all the
particles are more mobile and the particles in the veins move in
opposite directions.
Further analysis, that goes beyond the goals of the present work, is
necessary to understand the effect of the vicinity of the solvophilic
wall and if 
the veins are related to point-like
defects as seems to be suggested by Fig.\ref{fig:config_t_z2-z3}. 

%

\begin{figure}[H]
\begin{center}
\hspace*{-7cm}\includegraphics[width=6cm]{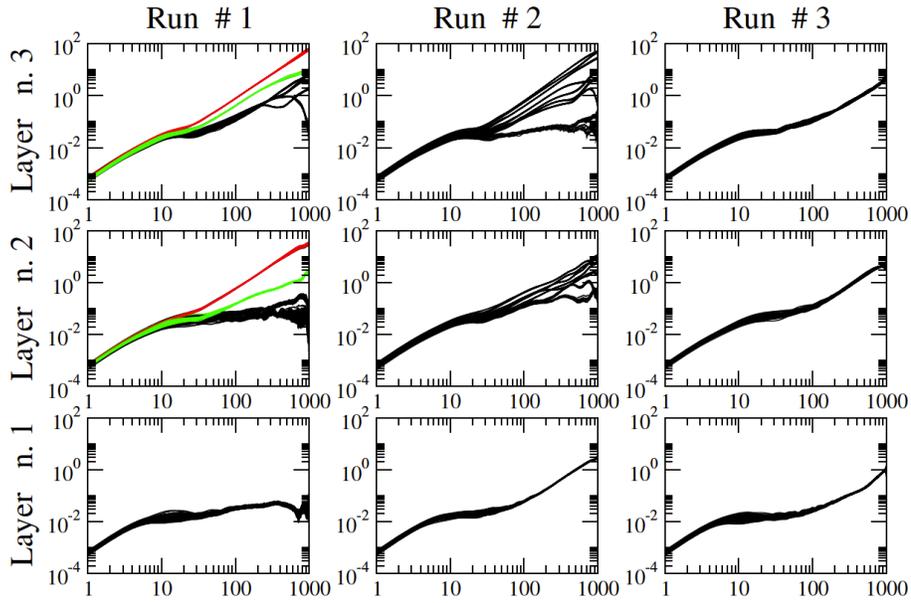} 
\end{center}
\vspace{0.3cm}
\caption{\label{fig:diff_z1-2-3_rand1-2-3} Single-particle MSD for
  layers and runs that are in Fig.\ref{fig:snap_z1-2-3_rand1-2-3}. For
  Run \#1, red and green colors are used for those particles belonging
  to veins performing a dynamics different from the rest of the
  particles in the layer.} 
\end{figure}
%

\begin{figure}[H]
\begin{center}
\vspace*{4cm}
\hspace*{-3.5cm}\includegraphics[width=2.5cm]{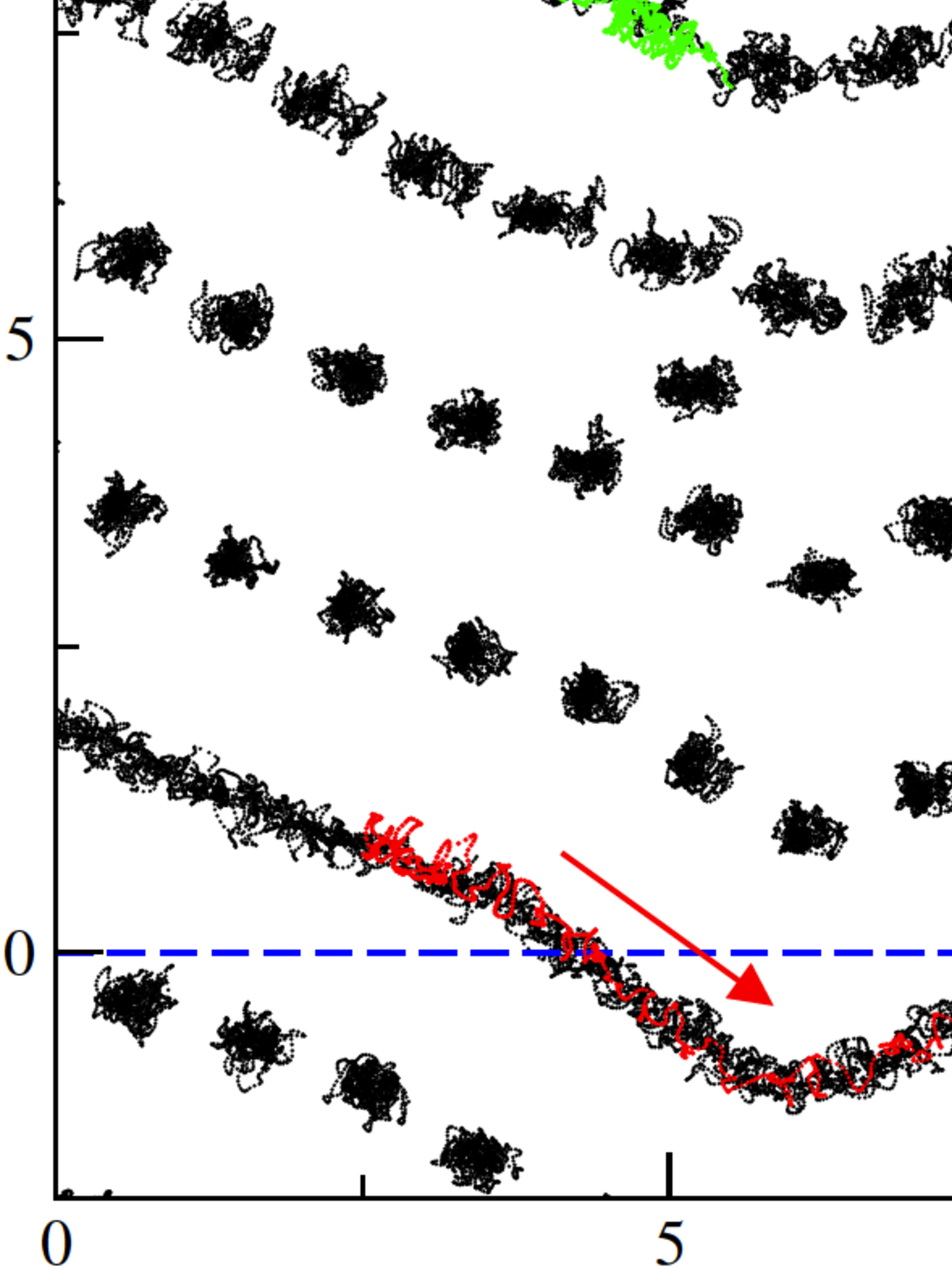}\hspace{5cm}
\includegraphics[width=2.5cm]{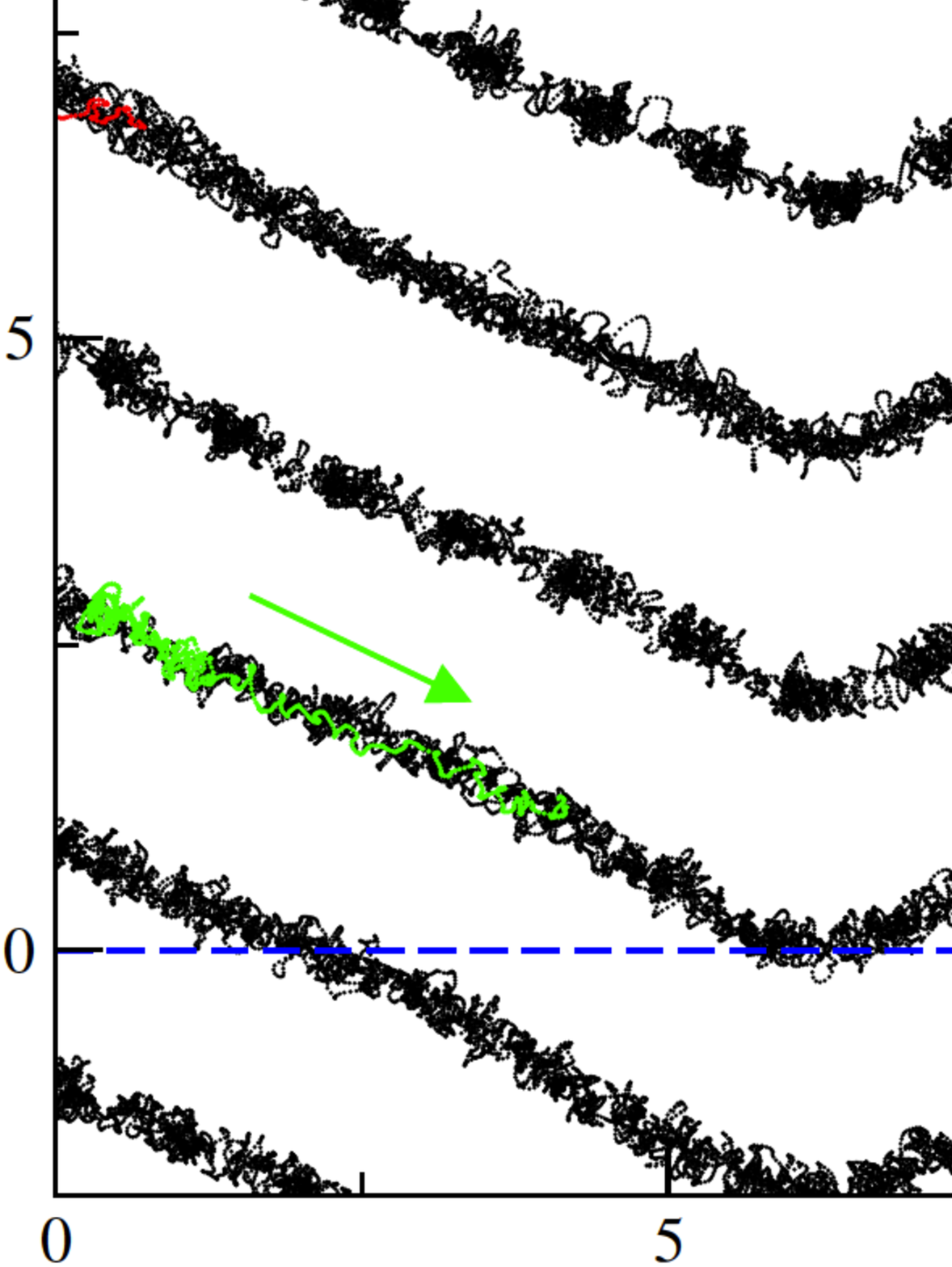}
\end{center}
\vspace{0.3cm}
\caption{\label{fig:config_t_z2-z3} In-layer trajectories for
  particles in the second (a) and third (b) layer of Run \#1 in
  Fig.\ref{fig:snap_z1-2-3_rand1-2-3}. In red and green we show the
  trajectories of representative particles belonging to stripes
  veins. The MSD of these particles is represented with the same color
  code in Fig.\ref{fig:diff_z1-2-3_rand1-2-3}. We apply periodic
  boundary conditions for $y^*=0,L_y$ where $L_y\simeq 12.1$ (dashed
  lines) and for $x^*=0,L_x$ where $L_x=15$.}   
\end{figure}


\subsection{Voronoi tessellation and structural analysis of the solid
  state} 
\label{sec:structural}

Our MSD and SP analysis show that at low $T$ and high $\rho$ there are 
layers taht behave as a solid. However, by looking only at the MSD and
SP is not possible to establish if a solid layer is in an amorphous,
crystal or polycrystal state \cite{Gerbode:2010dq}.
In order to better understand the structure of solid layers, we
computed the standard 2d Voronoi tessellation (useful to identify
defects present in the crystal structures, as vacancies, Frenkel-like,
dislocations and grain boundaries), and a modified version of it
(suitable to identify distorted crystal structures). 
With this analysis we can also disentagle the role that the three
relevant length scales (the diameter of the particles $a$, the
repulsive radius $R_R$, and the attractive minimum $R_A$), giving rise
to two competing length scale $R_R/a$ and $R_A/a$, play in the
determination of layer's structure. 
Indeed, the interdistance between two adjacent layers is $\sim R_R$,
while when stripes form in a specific layer for intermediate
densities, just to consider a specific case, particles within a stripe
are compressed at a distance $\sim a$, while the distance between
stripes depends on $\sim R_R$ and $\sim R_A$, as discussed in the
last part of this section. 

In the standard Voronoi tessellation we construct polygons centered
around particles forming a lattice whose edges are crossed in their
middle point by the edges of the Voronoi cells. This procedure
garantees that each Voronoi cell represent the proper volume of each
particle. To better visualize the result, we represent Voronoi cells
having a different number of edges with different colors. 
To reduce the noise in our analysis we adopt also a modified version
of the Voronoi tessellation in which we associate a color to a 
polygon in according to the number of edges of the polygon that have a 
length $10\%$ larger than that of the average edge lenght calculated
over the specific polygon itself. This procedure allows us to 
better visualize polycrystal structures despite the presence of small
lattice deformations.

\begin{figure}[H]
\begin{center}
\begin{minipage}{11cm}
\centering
\raisebox{-0.1\height}{\includegraphics[width=10.5cm]{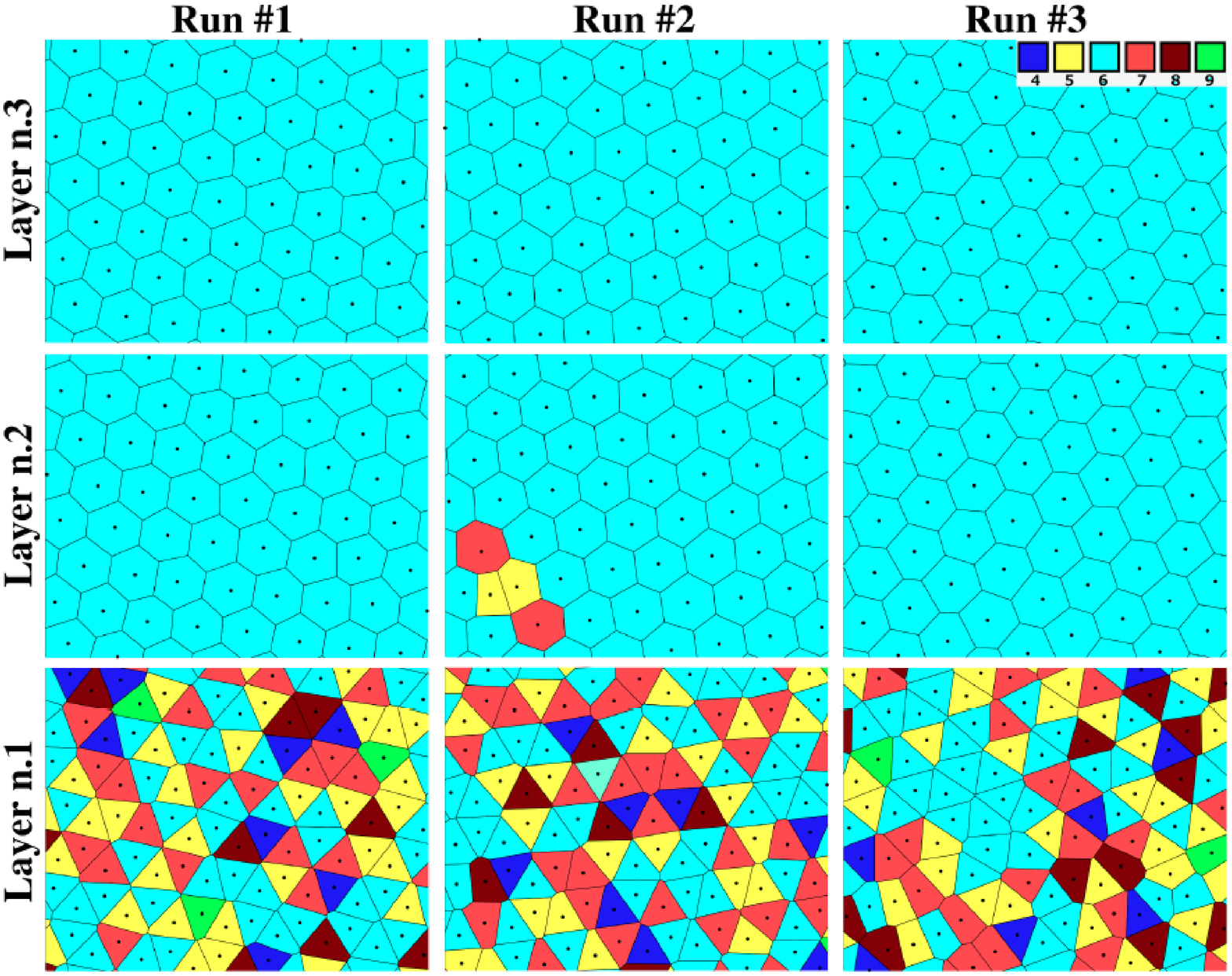}} 
\raisebox{-0.1\height}{\includegraphics[width=10.5cm]{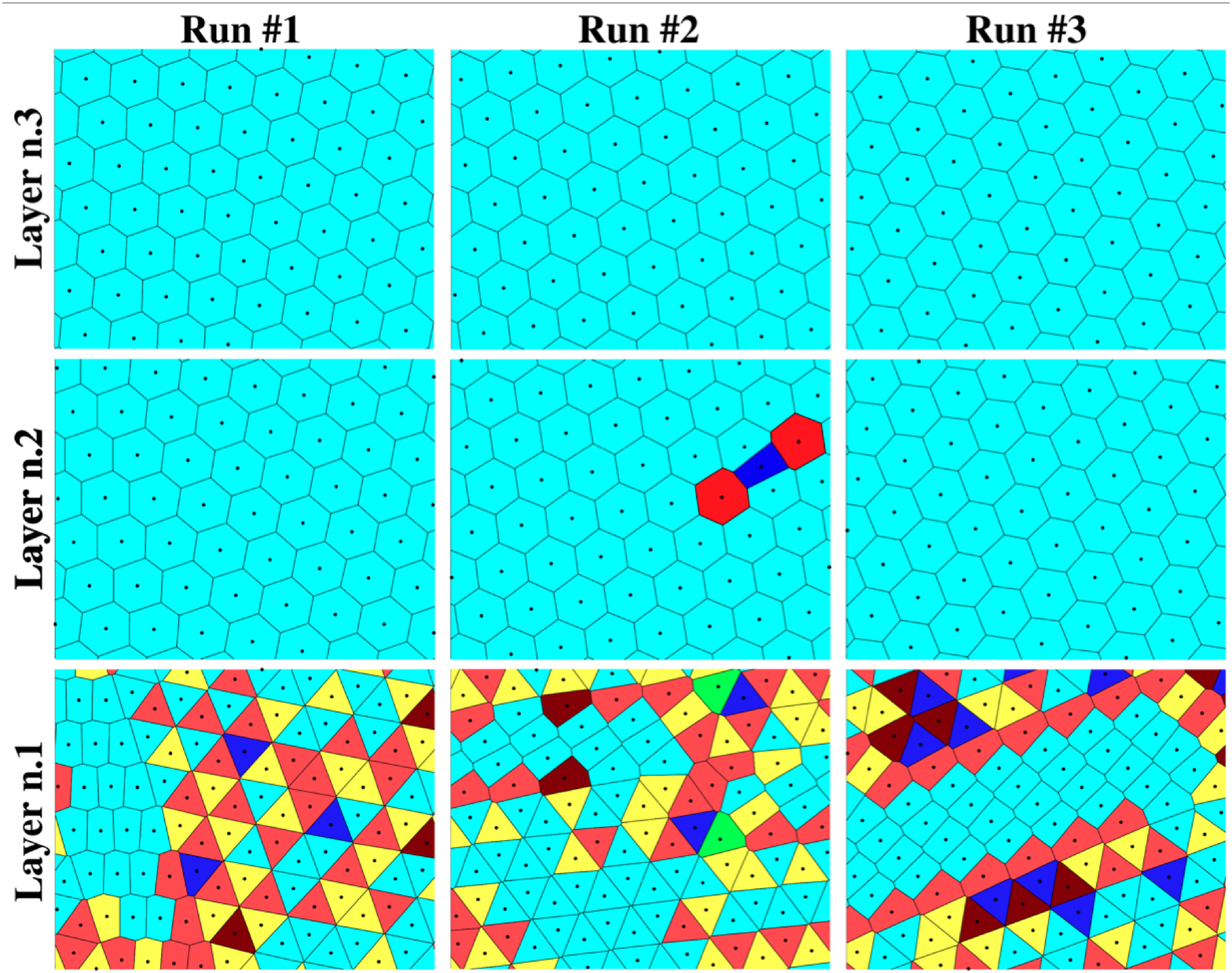}}
\end{minipage}
\end{center}
\vspace{0.3cm}
\caption{\label{fig:voronoi_rho0.11} 2d Voronoi tessellation for 
  $\rho^*=0.11$ and $T^*=0.3$ (top), and $T^*=0.0005$ (bottom). Only
  the first three layers near the bottom wall are displayed for three
  different runs. In the top-right part of the figure the color coding
  is reported.} 
\end{figure}

\begin{figure}[H]
\begin{center}
\begin{minipage}{11cm}
\centering
\raisebox{-0.1\height}{\includegraphics[width=10.5cm]{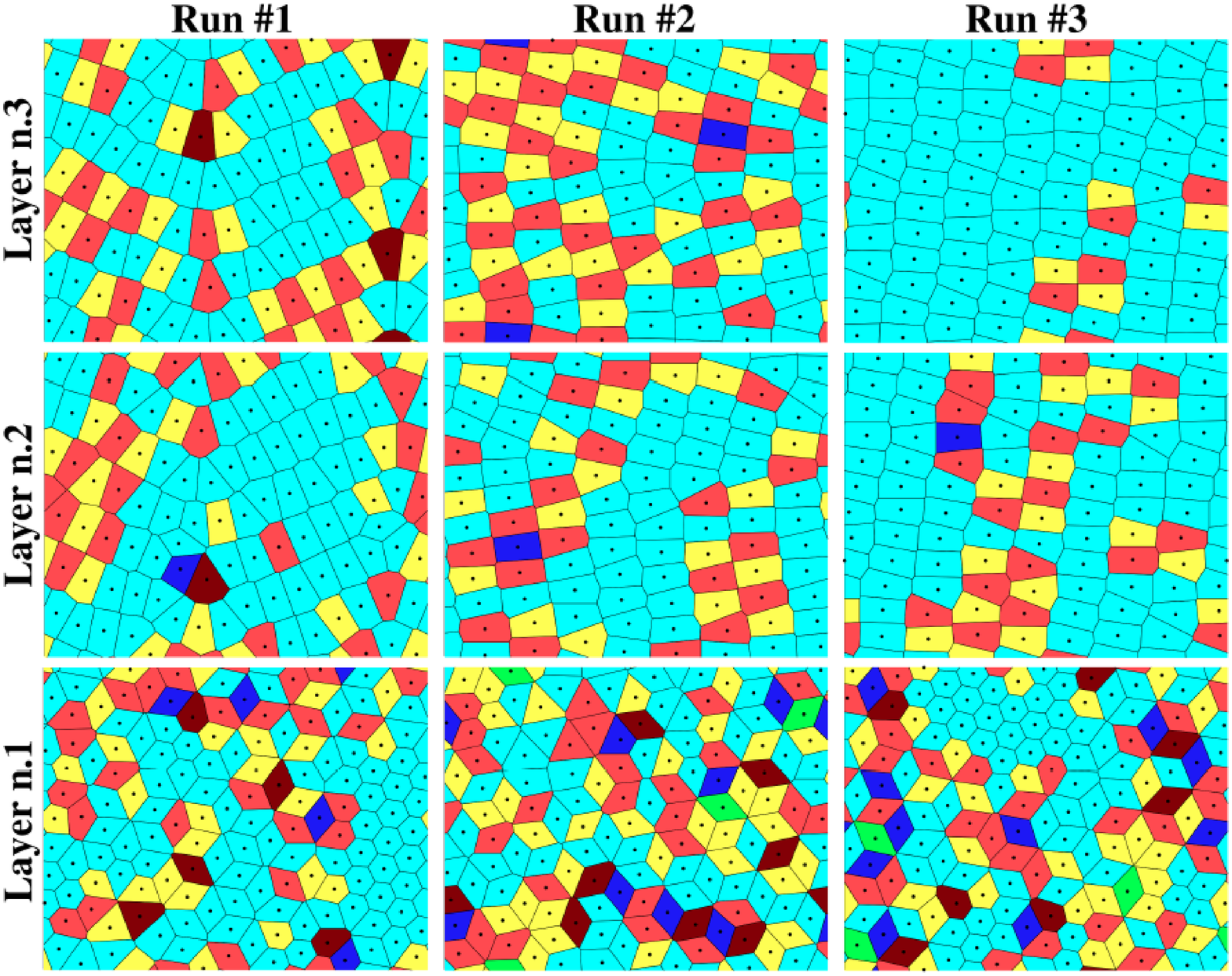}}
\raisebox{-0.1\height}{\includegraphics[width=10.5cm]{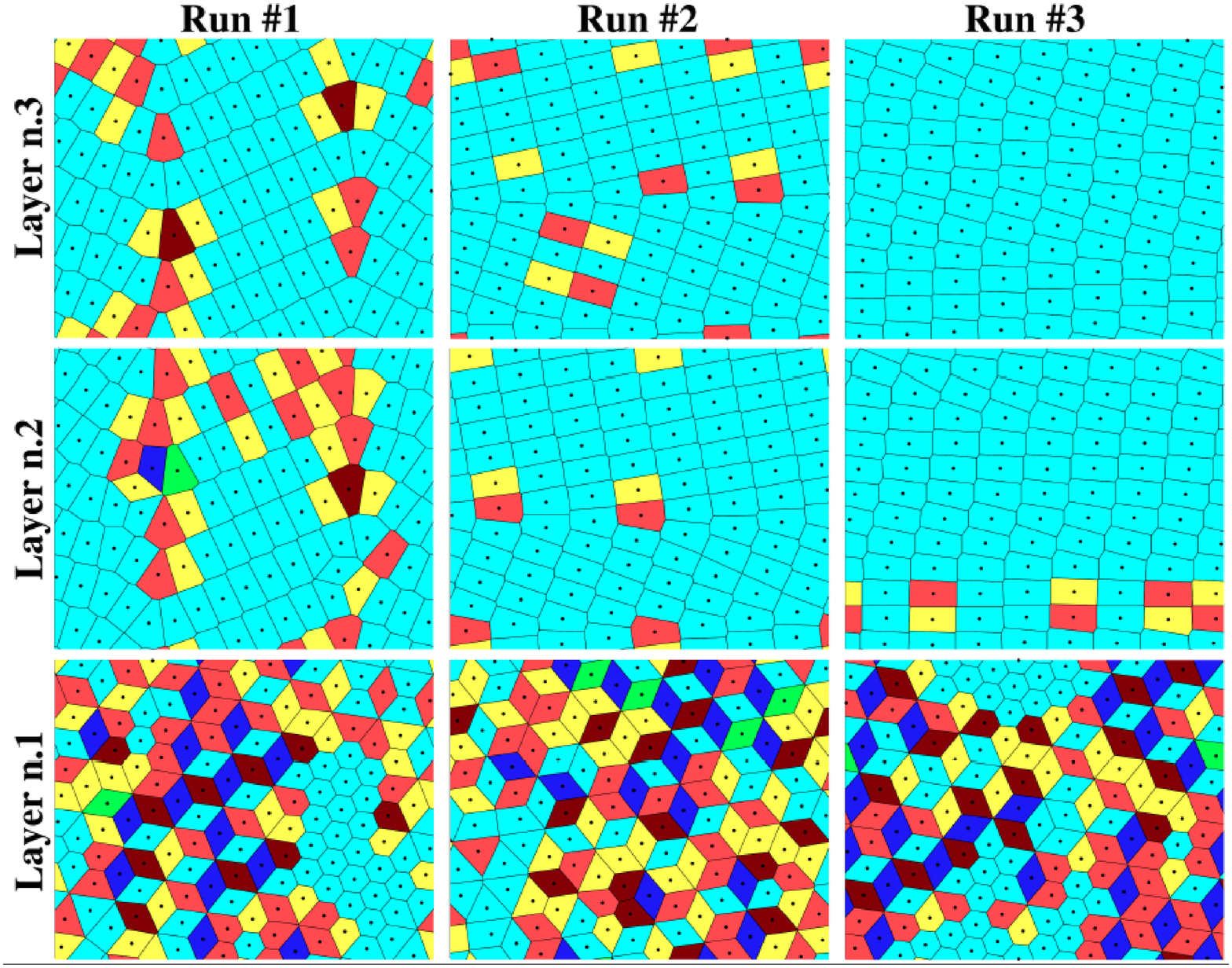}}
\end{minipage}
\end{center}
\vspace{0.3cm}
\caption{\label{fig:voronoi_rho0.30} 2d Voronoi tessellation for
  $\rho^*=0.30$ and $T^*=0.3$ (top), and $T^*=0.0005$ (bottom). Only
  the first three layers near the bottom wall are displayed for three
  different runs.} 
\end{figure}

We compute the Voronoi tessellation for low density ($\rho^*=0.11$)
and high density ($\rho^*=0.30$) at low temperature ($T^*=0.3$)
and very low temperature ($T^*=0.0005$), for three different
realizations of initial conditions (Figs.\ref{fig:voronoi_rho0.11}a,b,
\ref{fig:voronoi_rho0.30}a,b, and
Figs.\ref{fig:voronoi_rho0.11_mod2}a,b,
\ref{fig:voronoi_rho0.30_mod2}a,b in supplementary material).   
The configurations at temperature $T^*=0.0005$ are obtained by
annealing configurations equilibrated at $T^*=0.3$ with an annealing
rate of $0.025 U_A^{3/2}/(k_B a m^{1/2})$.

At low density ($\rho^*=0.11$), the first layer at $T^*=0.3$
(Fig.\ref{fig:voronoi_rho0.11_mod2}) is in a frustrated solid state
that by annealing toward $T^*=0.0005$
(Fig.\ref{fig:voronoi_rho0.30_mod2}) becomes a frustrated
polycrystal. The very low-$T$ polycrystal has point and line defects as 
grain boundaries dividing a deformed honeycomb lattice (the deformed
green triangles) from a stripe phase (the stretched hexagonal cyan
polygons). For both considered $T^*$ the other layers are organized in
a triangular lattice (where each particle is surrounded by a hexagonal
cyan polygon). We only observe defects, such as dislocations (Run \# 2
in Fig.\ref{fig:voronoi_rho0.11_mod2}a and
Fig.\ref{fig:voronoi_rho0.11}a) that are not eliminated by annealing
(Run \# 2 in Fig.\ref{fig:voronoi_rho0.11_mod2}b). 

At high density ($\rho^*=0.30$), the first layer at $T^*=0.3$
(Fig.\ref{fig:voronoi_rho0.30_mod2}a) is in a polycrystal state with
defects. At $T^*=0.0005$ (Fig.\ref{fig:voronoi_rho0.30_mod2}b) we
observe two principal crystal grains: a triangular lattice (cyan
polygons) and a Kagome lattice with defects (blue rhombouses).
At $T^*=0.3$, the layers $n=2,3$ present a zigzagging stripe structure
with orientation and angles that can change from run to run. At
$T^*=0.0005$ the stripe structure of these layers becomes more regular.
These observations emphasize that the increase of density induces an
increase of disorder in the solid layers, propagating from the layer
$n=1$ to the other layers and up to the layer $n=11$.
The formation of crystal defects during the annealing, and the fact
that they are different for different initial conditions, indicate
that the system cannot reach easely the global minimum of the free
energy landscape, corresponding to the crystal configuration, but is
trapped in local minima due to the slowing down of the dynamics and
the templating effect of the solvophilic wall. In particular, the
mismatch of the wall structure with the bulk crystal structure induces
a frustrating effect that is more evident near the wall (in layers
$n=1$ and $n=2$) for increasing density.

\begin{figure}[H]
\begin{center}
\begin{minipage}{11cm}
\centering
\raisebox{-0.1\height}{\includegraphics[width=10.5cm]{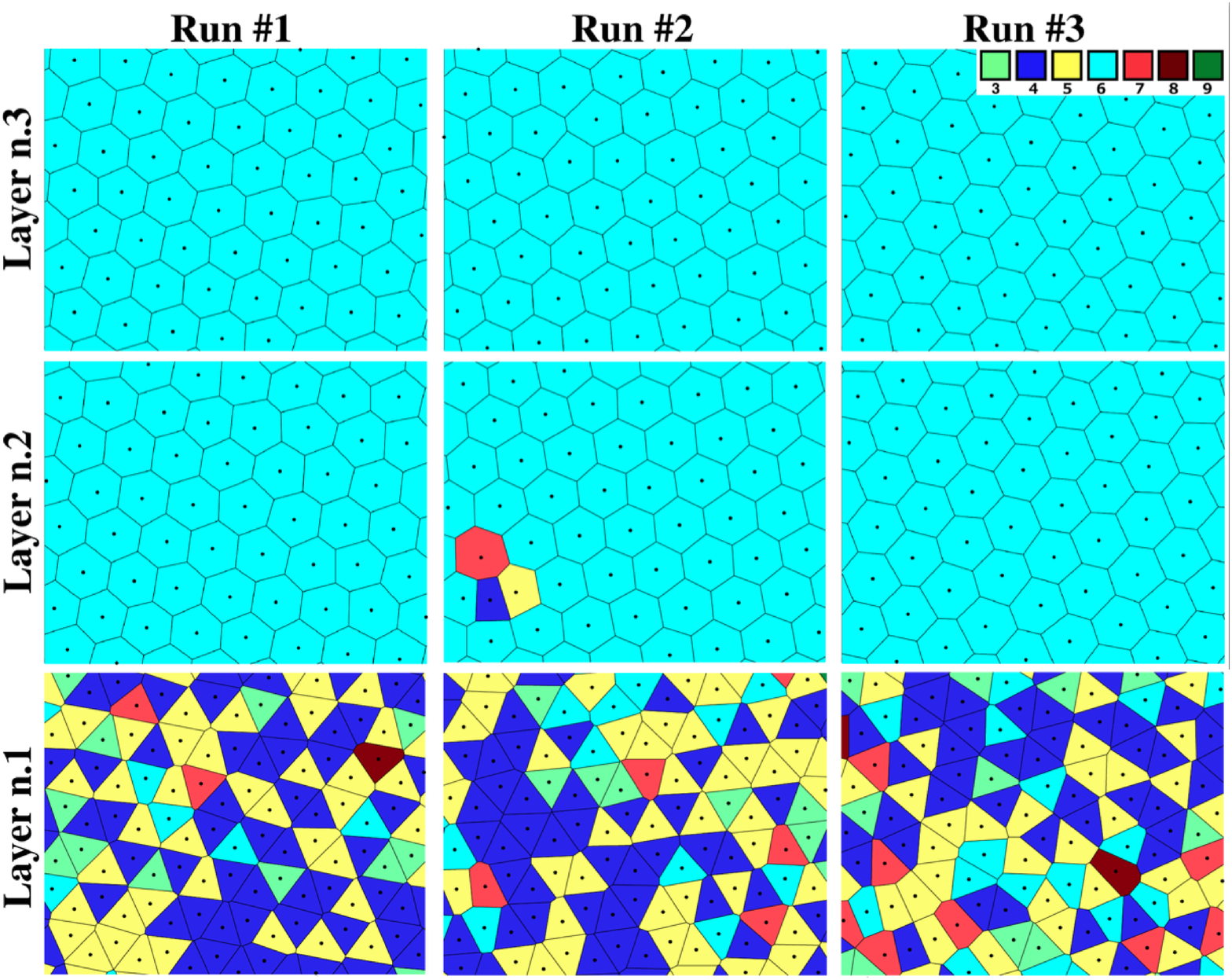}} 
\raisebox{-0.1\height}{\includegraphics[width=10.5cm]{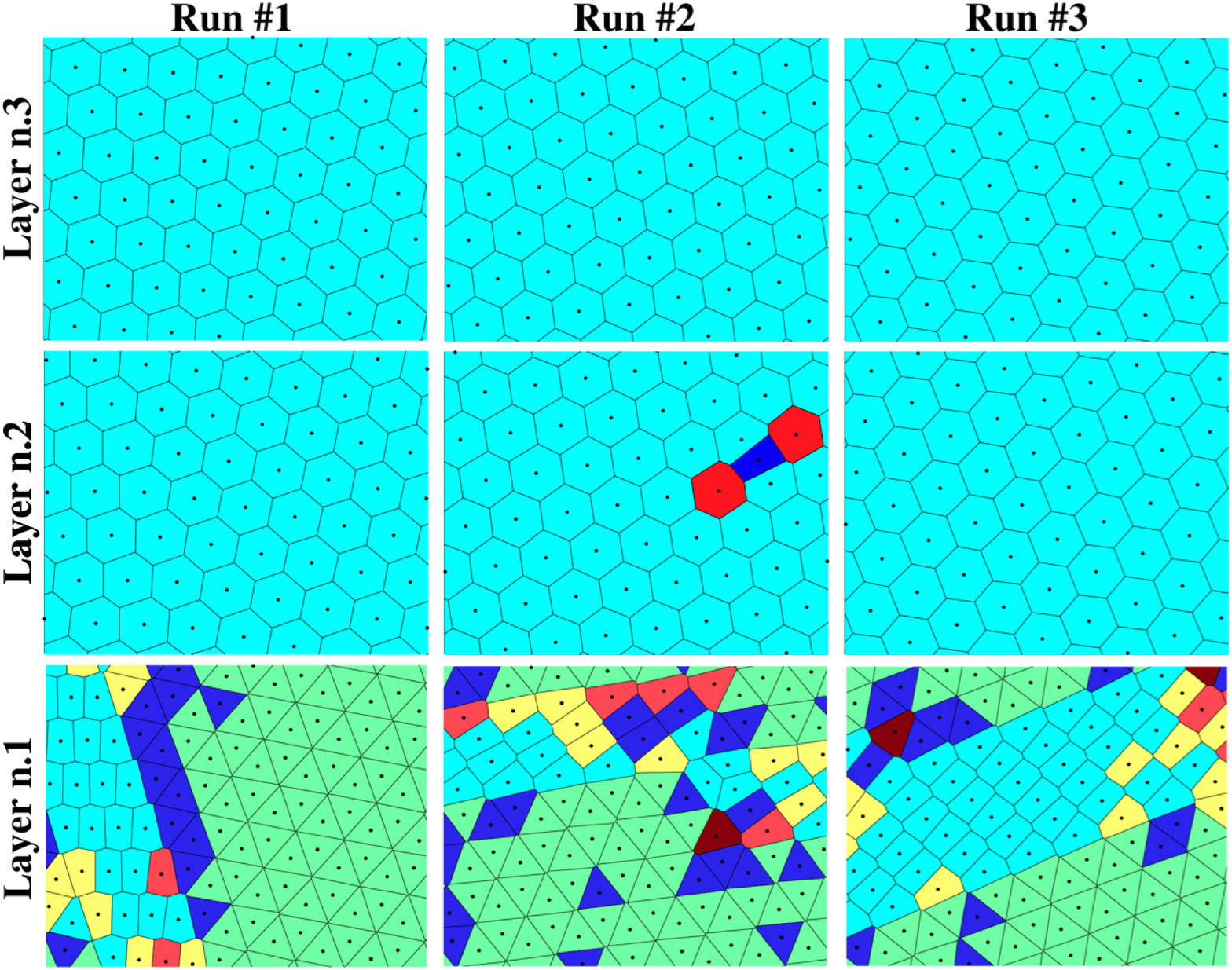}}
\end{minipage}
\end{center}
\vspace{0.3cm}
\caption{\label{fig:voronoi_rho0.11_mod2} Modified version of the 2d
  Voronoi tessellation, in according to the rule specified in the text, for 
  $\rho^*=0.11$ and $T^*=0.3$ (top), and $T^*=0.0005$ (bottom). We
  display the first three layers near the solvophilic wall for three 
  different runs. In the top-right part of the figure the color coding
  is reported as a function of number of edges of each Voronoi polygon.} 
\end{figure}

\begin{figure}[H]
\begin{center}
\begin{minipage}{11cm}
\centering
\raisebox{-0.1\height}{\includegraphics[width=10.5cm]{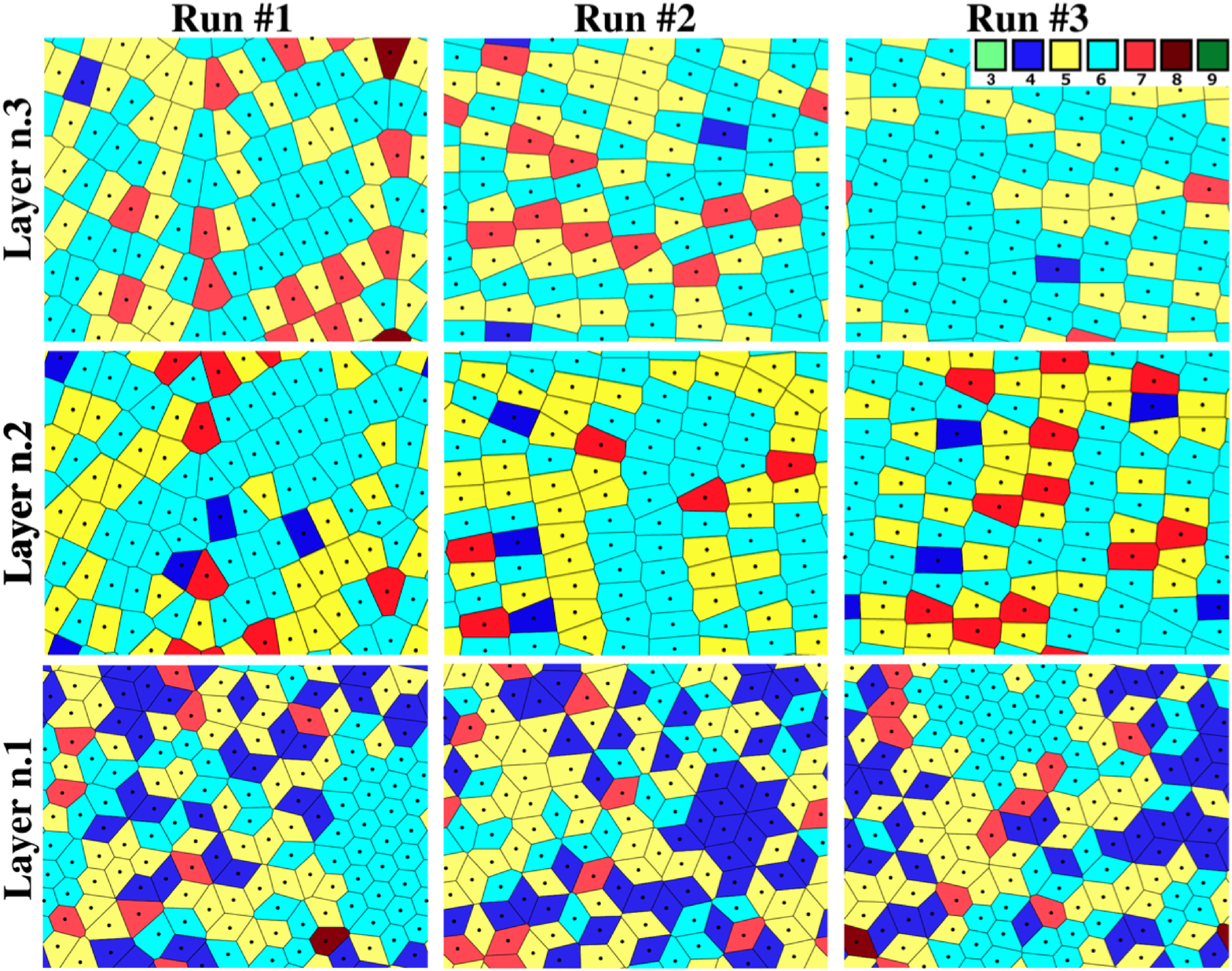}}
\raisebox{-0.1\height}{\includegraphics[width=10.5cm]{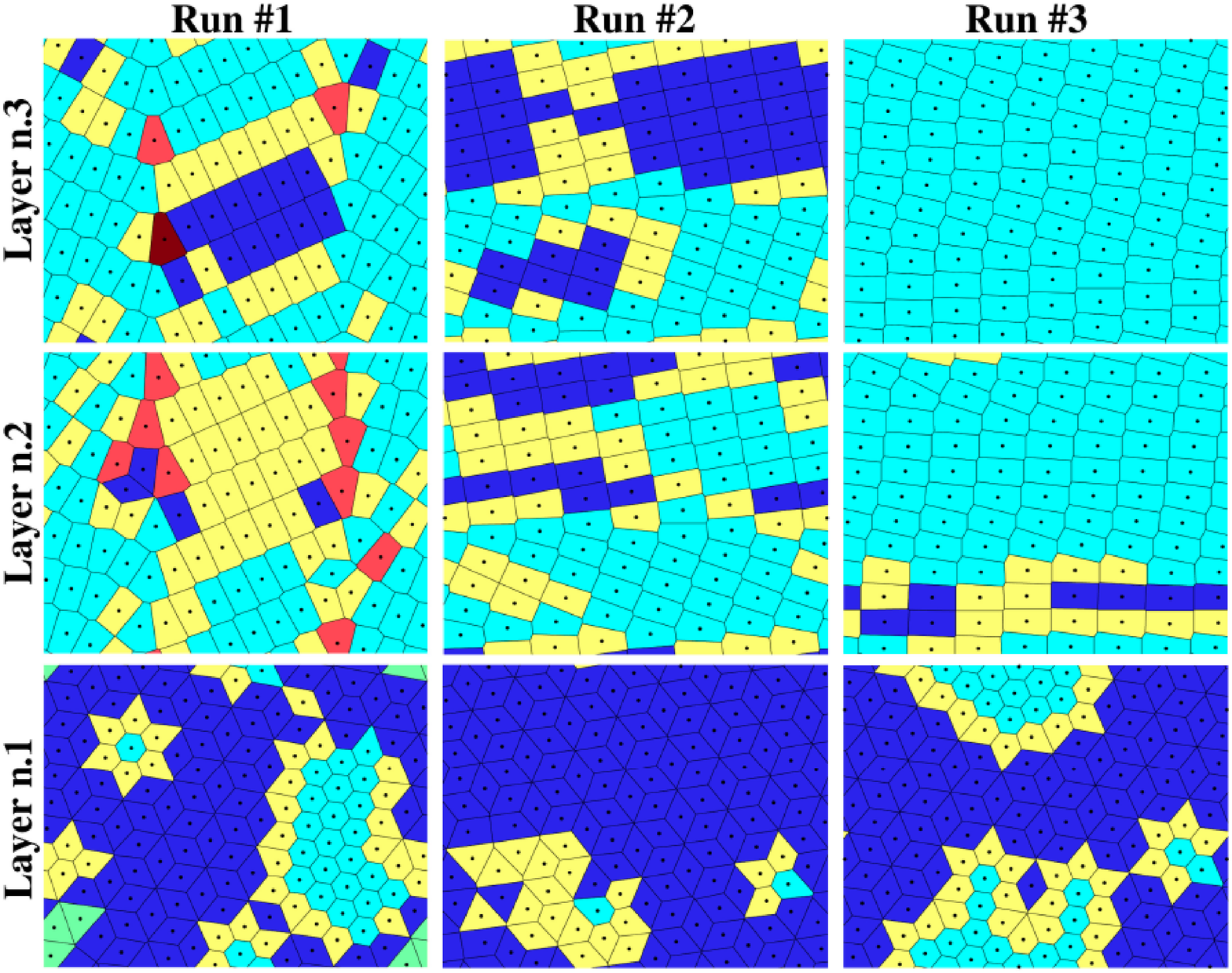}}
\end{minipage}
\end{center}
\vspace{0.3cm}
\caption{\label{fig:voronoi_rho0.30_mod2} As in
  Fig.\ref{fig:voronoi_rho0.11_mod2}, but for $\rho^*=0.30$ and
  $T^*=0.3$ (top), and $T^*=0.0005$ (bottom). For the layer $n=1$ the
  triangular lattice (cyan hexagons) and the Kagome lattice (blue
  rhombouses) are clearly visible.}
\end{figure}

To understand the formation of stripes, we follow the same approach as
in Refs~\cite{Han:2008ly, shokef2011} to show that, if the principal
contribution to the minimization of the free energy comes from the
energetic therm, in the discreet potential approximation, under
suitable conditions of density and temperature, particles organize in
straight or zigzagging stripes.
In the discreet potential approximation the energetic terms can be
reduced to the soft core ($U_R$) and the attractive well ($U_A$). As a 
consequence the energetic contribution coming from the
interaction with particles of the adjacent layers is 
approximately  constant when the layer density is fixed. 
Therefore, the energetic cost of stripes formation is determined only
by the contribution of the in-layer particle interactions.  
In particular, if a layer has a triangular structure, as the
stable configuration of layers $n=$2, ...,11 at low density
(Fig.\ref{fig:voronoi_rho0.11_mod2}a,b), then for
sufficiently high density the layer will prefer to form stripes. 
\begin{figure}[H]
\begin{center}
\includegraphics[width=7cm]{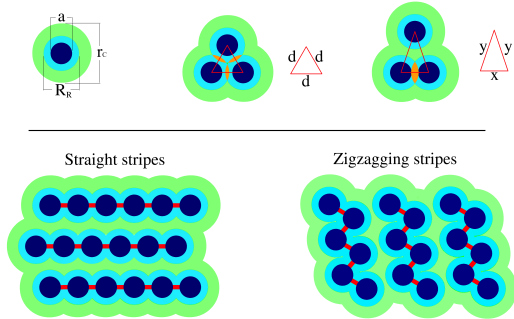}
\end{center}
\vspace{0.3cm}
\caption{\label{fig:stripes} 2d schematic representation of particles
  composed by an hard core (in dark blue) of size $a$, a soft corona (in
  cyan) of size $R_R$ and an attractive external corona (in green)
  that extends uo to the potential cutoff $r_C$. The equilateral and
  isosceles triangles represent the unitary cell of the triangular and
  straight-stripes lattice, respectively. An example of (maximal)
  zigzagging-stripes of the same density of straight-stripes is also
  shown.} 
\end{figure}

Consider our fluid made of particles with a hard core surrounded by a
soft corona and an external attractive corona
(Fig.\ref{fig:stripes}). If $d$ is the lattice constant of the
triangular structure at the soft-corona distance, then the density
of the layer is $\rho_l=N_l/(L_xL_y)=2/(\sqrt{3}d^2)$, where $N_l$ is
the number of particles present in this layer. 
If we allow the triangular lattice to deform in order to minimize the
energy of the layer, the new unit cell will be composed by the
isosceles triangle in which one side is equal to $x$ and the other two
are equals to $y$ with $x<d<y$.  
The fact that the density doesn't change implies that 
$y=\sqrt{(3d^4)/(4x^2)+x^2/4}=\sqrt{1/(\rho^2x^2)+x^2/4}=y(\rho_l,x)$. 

The energy per particle of the layer is
$U_l=[(\sqrt{N_l}-1)/\sqrt{N_l}]U_x+2[(\sqrt{N_l}-1)/\sqrt{N_l}]U_y$.
For $N_l\gg 1$ it becomes $U_l\simeq U_x+2U_y$, where
$U_x$ and $U_y$ are the energy associated to the interaction between
the particle along the $x$ and $y$ side of the triangle respectively.

In the discreet potential approximation it is $U_x=U_R$ for $a<x<R_R$,
$U_x=-U_A=-U_R/2$ for $R_R<x<r_C$ and $U_x=0$ for $x>r_C$ (note that
we obtain the same result if instead of $r_C$ we consider any value
between $R_R$ and $r_C$. Indeed, the present approach has been applyed
to show the stability of stripes cnofiguration for a pure repulsive
potential model \cite{Han:2008ly, shokef2011}). The same holds for
$U_y$ substituting $x$ with $y$.       

For sufficiently high density, i.e. for $d<R_R$ or
$\rho_l>2/(\sqrt{3}R_R^2)$, the energies per 
particle associated to a layer formed by equilateral or isosceles
triangles are $U^{equi}_l=3U_R$ or $U^{iso}_l=0$, respectively. 
Therefore, under these conditions, the layer will prefer to form
stripes. 
Furthermore, from geometric consideration it is possible to conclude
that the zigzagging-stripe lattice can be obtained as a deformation of
the straight-stripe lattice without changing the density and keeping the
energies per particle $U^{iso}_l\simeq 0$ \cite{Han:2008ly, shokef2011}.
In general, many different zigzagging stripe lattices are possible all
with comparable energy per particle (Fig.\ref{fig:stripes}).

Considering the stripes that form in the layers $n=2,...,9$ for system
density $\rho^*=0.30$ and temperature $T^*=0.3$, the resulting 
average in-layer density is $\rho_l\simeq 0.53$, and $x\simeq
1.05$. Hence, from the previous equation for $y=y(\rho_l,x)$, 
we find $y\simeq 1.87$. In view of all the approximation made, we
consider this value  consistent
with $y\simeq2$ of the distance
between the closest particles belonging to two adjacent stripes in the same
layer.


\section{Conclusions}
\label{sec:conclusions}

By considering many layers of a confined anomalous fluid 
  \cite{1359736,2216695,2013arXiv1307.3238B,Krott2013} we 
show that the effect of the structured solvophilic wall can extend up
to the entire slit pore.
In particular, we study structural and dynamical properties of a
monocomponent anomalous liquid under confinement. The fluid has two
characteristic distances and can be considered as a coarse-grained
model for globular proteins \cite{C3SM50220A}, colloidal systems
\cite{Lowen1997129,PhysRevE.55.637,PhysRevLett.74.2519} or,
to some extent, liquid metals with water-like anomalies
\cite{Bryk:2013fk}.  
We perform molecular dynamics simulations of the fluid in a slit pore
with a solvophilic wall and a solvophobic wall. The solvophilic wall
has structure while the solvophobic one has no structure. We observe that
the molecules organize in an inhomogeneous way, forming layers that
are parallel to the surfaces, with higher density near the solvophilic
surface with respect to the center of the slit pore. For sufficiently
high densities, for which the fluid occupy entirely the pore, we
observe an increase of density also close to the solvophobic surface,
but in a less prominent way. These results are consistent with
experimental and theoretical works for nanoconfined fluids. 

At low
temperature we observe coexistence between the homogeneous liquid,
heterogeneous liquid and solid phase of the fluid. The influence of
the structured solvophilic surface on the solid layers can extend as
far as the sixth hydration layer at low $T$ and high $\rho$. In
particular, we find a strong correlation between the structure of the
solvophilic surface and that of the first layer suggesting a
``templating'' effect. Indeed, the large density of solvophilic
surface particles allows the formation of a first layer at high
density. The high 
energy cost of this first layer is compensated by the large number of attractive
interactions between the particles of the first layer and those
of the surface. Further energy gain comes from an extra energy
term due to the first in-layer particle interactions.  

Moving further from the wall, we find that 
the first layer has a ``molding'' effect on the second
layer. This is because the  density of the first layer is  smaller
than that of the surface and is not high-enough to propagate its
template. Nevertheless, the low-density second layer is in condition
to template the third layer replicating its structure and inducing a
long-range effect that can, eventually, involve the whole system.

From the calculation of the mean square displacement and the Voronoi
tessellation we conclude that at low temperature the first layer close
to the solvophilic surface is a polycrystal with two competing phases
that generate low-energy states with high degeneracy and very slow
dynamics. 
At low densities the two competing phases are stripes and honeycomb
lattice, while at high densities are triangular and kagome lattice. We
understand this result as a consequence of the high density
(triangular) structure of the solvophilic wall with a lattice step
that corresponds to the hard repulsive distance of the solvent, and
the strong wall-solvent attractive interaction.
These properties of the wall generate in the first solvent layer local
regions with density and energy that are higher than the average of
the layer. As a consequence, other regions within the layer have
density and energy below the average, giving rise to a competing
crystal structure.
Our results remind us of some recent experiments and simulations for a
thin-film of water on $\mbox{BaF}_2(111)$ surface for which the
authors found a very high density first interfacial layer for all
temperatures, while they would expect, from thermodynamic arguments, a
lower density liquid at supercooled conditions \cite{Kaya2013}. In
other recent experiments and simulations \cite{Chen:2011ve,
  Cox:2012vn}, the authors pointed out that it is necessary to revise
the theory of heterogeneous nucleation when the crystalization induces
a non-zero entropy at zero temperature and the system initially is far
from equilibrium. 

Apart from the layers close to the two surfaces, we observe that the
structure of each layer mainly depends on its density. In the case of
the stripe phase, using simple geometrical and energetic
considerations, we find that straight and zigzagging stripes are the
stable configurations for intermediate densities.

Furtheremore, analysing the mean square displacement layer by layer,
we observe layers at high densities and low temperatures with a
caging-like behavior characterized by a ballistic dynamics followed by
an arrested state (plateau) and a superdiffusive regime with a
diffusion exponent $1<\alpha<2$. Our analysis shows that this
behavior is due to the formation of liquid veins within the stripe
phase. In particular we observe that each vein can behave differently
from the others diffusing in one of the two possible directions along
the stripes and having a different diffusion exponent $1<\alpha<2$.
We rationalize the different possible values of $\alpha$ as a
consequence of the presence of residual stress that could introduce an
effective force acting on the fluid. Under suitable conditions, e.g. a
constant effective force along the stripe, the particles in the vein
could perform a biased one-dimensional random walk characterized by an
exponent of the MSD that approaches the ballistic value ($\alpha=2$).
The behavior of these veins can be analyzed in a more quantitative way
by computing, for example, the temporal autocorrelation function, the
intermediate scattering function
\cite{Milischuk2012,Ingebrigtsen2013}, the relative displacement of
nearest neighbors or the particles displacements following the
Lindemann criterion \cite{Zahn:1999uq}. We will present this analysis
in future works. 

Our results show that the dynamical slowing down of the anomalous
solvent near the solvophilic wall does not imply by necessity the
complete freezing of the first hydration layers, because at low $T$
and high $\rho$ we observe largely heterogeneous dynamics in three
layers with the formation of liquid veins within a frozen matrix of
solvent. Therefore, under these considerations the partial freezing of
the first hydration layer does not correspond necessarily to an
effective reduction of the channel section in terms of transport
properties, at variance with the conclusions of
Ref.\cite{doi:10.1021/jp307900q}.

\section*{Acknowledgments}
We thank Y. Shokef for useful discussion on stripes formation.

\section{Supplementary information}
\label{sec:supplementary}

\end{document}